\documentstyle[aps,epsfig,float,amstex]{revtex}
\newcommand{\be}{\begin{equation}}
\newcommand{\ee}{\end{equation}}
\newcommand{\beq}{\begin{eqnarray}}
\newcommand{\eeq}{\end{eqnarray}}
\newcommand{\bi}{\bibitem} 
\begin{document}
    
\def\gC{\mbox{\boldmath $C$}}
\def\gZ{\mbox{\boldmath $Z$}}
\def\gR{\mbox{\boldmath $R$}}
\def\gN{\mbox{\boldmath $N$}}
\def\ua{\uparrow}
\def\da{\downarrow}
\def\a{\alpha}
\def\b{\beta}
\def\g{\gamma}
\def\G{\Gamma}
\def\d{\delta}
\def\D{\Delta}
\def\e{\epsilon}
\def\ve{\varepsilon}
\def\z{\zeta}
\def\h{\eta}
\def\th{\theta}
\def\k{\kappa}
\def\l{\lambda}
\def\L{\Lambda}
\def\m{\mu}
\def\n{\nu}
\def\x{\xi}
\def\X{\Xi}
\def\p{\pi}
\def\P{\Pi}
\def\r{\rho}
\def\s{\sigma}
\def\S{\Sigma}
\def\t{\tau}
\def\f{\phi}
\def\vf{\varphi}
\def\F{\Phi}
\def\c{\chi}
\def\w{\omega}
\def\W{\Omega}
\def\Q{\Psi}
\def\q{\psi}
\def\de{\partial}
\def\inf{\infty}
\def\ra{\rightarrow}
\def\bra{\langle}
\def\ket{\rangle}
\title{Antiferromagnetism in the Exact Ground State of the Half Filled 
 Hubbard Model on the Complete-Bipartite Graph}

\author{Gianluca Stefanucci and Michele Cini}
\address{Istituto Nazionale per la Fisica della Materia, Dipartimento di Fisica,\\
Universita' di Roma Tor Vergata, Via della Ricerca Scientifica, 1-00133\\
Roma, Italy}
\maketitle

\begin{abstract}

As a prototype model of antiferromagnetism, we propose a repulsive Hubbard
Hamiltonian defined on a graph $\L={\cal A}\cup{\cal B}$ with ${\cal A}\cap
{\cal B}=\emptyset$ and bonds connecting any element of ${\cal A}$ with all 
the elements of ${\cal B}$. Since all the hopping matrix elements associated
with each bond are equal, the model is invariant under an arbitrary 
permutation of the ${\cal A}$-sites and/or of the ${\cal B}$-sites. This is 
the Hubbard model defined on the so called 
$(N_{A},N_{B})$-complete-bipartite graph, $N_{A}$ ($N_{B}$) being the number of 
elements in ${\cal A}$ (${\cal B}$). In this paper we analytically find the 
{\it exact} ground state for $N_{A}=N_{B}=N$ at half 
filling for any $N$; the ground state expectation value of the repulsion 
term has a  maximum  at a
critical $N$-dependent value of the on-site Hubbard $U$ and then drops 
like $1/U$ for large $U$.
The wave function and the energy of the unique, singlet ground state assume 
a particularly elegant form for $N \ra \inf$.  We also calculate the spin-spin 
correlation function and show that the ground state exhibits an antiferromagnetic 
order for any non-zero $U$ even in the thermodynamic limit. 
This is the first explicit analytic example of  an antiferromagnetic ground 
state in a Hubbard-like model of itinerant electrons. The kinetic term 
induces non-trivial correlations among the particles and an antiparallel 
spin configuration in the two sublattices comes to be energetically  
favoured at zero Temperature. On the other hand, if the thermodynamic 
limit is taken and then zero Temperature is approached, a paramagnetic 
behavior results. The thermodynamic limit does not commute with the 
zero-Temperature limit, and this fact can be made explicit by the 
analytic solutions.

\end{abstract}
  
\bigskip
{\small 

\section{Introduction}

The Hubbard Hamiltonian is one of the most popular models to describe
strongly correlated electron systems. In spite of its simple 
definition, it can be exactly solved only in few cases. 
An example is the milestone solution of the Hubbard 
ring\cite{liebwu} by means of the  Bethe-{\it Ansatz}\cite{bethe} 
extended to fermions\cite{fermbethe}\cite{yang2}. 
Other exact solutions were worked out for the one-dimensional 
Hubbard ring in the presence of 
an external magnetic field\cite{suzhao}, for the one-dimensional 
Hubbard chain with open boundary conditions\cite{ligruber} and 
thereafter for the 
$SU(N)$ one-dimensional Hubbard ring\cite{houpeng}. However, when the 
space dimensionality becomes bigger than 1, few exact results are 
available and usually they concern the ground state properties. Among 
them, we mention the Lieb theorem on the ground state spin 
degeneracy\cite{lieb} that will be explicitly used in this work. 
Exact ground state wave functions are even more infrequent. To the 
best of our knowledge the only non-trivial results are the 
ferromagnetic ground-state solutions devised by Nagaoka\cite{nagaoka} 
(infinite repulsion and one hole over the half filled 
system), by Mielke and Tasaki\cite{mielke}\cite{tasaki} 
(for which the kinetic energy spectrum is macroscopically 
degenerate or at least very flat) and by Wang\cite{wang}  
(infinite-range hopping). 

In the Hubbard Hamiltonian with infinite-range 
hopping a particle in any site can hop to any other site of the 
system; the associated graph is said to be {\em complete}. This model 
was numerically studied by Patterson on small clusters\cite{patterson} 
in 1972 and solved in the thermodynamic 
limit by van Dongen and Vollhardt\cite{vDV} only at the end of the 
eighties. Much more effort was needed to find the exact ground 
state(s) in the finite-size system. Verges {\it et al.} managed to 
accomplish this task for arbitrary numbers of particles and sites in 
the limit of infinite on-site repulsion $U$\cite{verges} by exploiting 
a scheme proposed by Brandt and Giesekus \cite{brandt}. Two years 
later Wang constructed explicitly the ground states of the system for arbitrary 
$U$ and number of particles above half filling\cite{wang}. In the 
case of one particle added over the half-filled system the 
ferromagnetic ground-state solution follows as a special (and the 
easiest) case of general results by Mielke and Tasaki\cite{mita}. 

In this paper we find the {\it exact} ground state wave function of the  
half-filled Hubbard model on the Complete-Bipartite Graph (CBG) for arbitrary 
but finite $U$. The CBG $\L={\cal A}\cup{\cal 
B}$ has bonds connecting any element of ${\cal A}$ with all 
the elements of ${\cal B}$ and can be considered as the natural further 
step (with respect to the complete graph previously described) 
towards the standard Hubbard Hamiltonian (defined on the hypercubic 
lattice and hopping between nearest-neighbours sites). Even if the CBG is still somewhat unrealistic, we feel that exact solutions should 
always be welcome, expecially because they lend themselves to be 
generalized. 
Furthermore, our solution is the first example of {\it antiferromagnetic} ground 
state in a model of itinerant electrons, in contrast with the 
ferromagnetic solutions mentioned above; it may provide  
useful hints about  antiferromagnetism  outside of the 
strong coupling regime (where the Hubbard model can be mapped onto 
the Heisenberg model). 

The paper has been organized as follows. 
In Section \ref{model} we define the model and discuss the physics of 
the non-interacting ($U=0$) Hamiltonian together with few  relevant 
examples of finite-size realizations.  
In Section \ref{thermolimi} we study the thermodynamic limit of a 
class of Hubbard-like models, which includes the Hubbard Hamiltonian 
on the complete graph and on the complete bipartite graph, having a 
non-extensive number of isolated one-particle energy levels plus a single 
level whose degeneracy is proportional to the size of the system. Following the 
reasoning of van Dongen and Vollhardt\cite{vDV}, we show that the kinetic term 
is totally decoupled from the interaction term and that the system behaves 
as a paramagnet: the spin-spin correlation length is zero. These  
results are a consequence of the trivial behaviour of such models any 
time the thermodynamic limit is taken first. On the contrary, much 
more difficult is to find exact properties in finite-size systems and 
different answers may be obtained if the thermodynamic limit is taken 
only at the end of the calculations (as we shall show in 
Section \ref{results}). In Section \ref{w=0states} we introduce the essential tools to 
face the problem. Let $N=N_{A}=N_{B}$ be the number of sites of each 
sublattice ${\cal A}$ and ${\cal B}$. First, we find a ($2N-2$)-body 
determinantal eigenstate $|\F_{AF}\ket$ of the Hamiltonian with vanishing double occupation; 
then, we demonstrate that it is a key tool to build the ground state. 
We show that mapping the 
${\cal A}$-sites onto the ${\cal B}$-sites and {\it viceversa}, $|\F_{AF}\ket$ retains 
its form except for a spin-flip; we shall call this property the {\it 
antiferromagnetic property}. Several analogies with the properties of 
the half-filled standard $N \times N$ Hubbard model are pointed out at 
this stage. Here we also deal with the spin projection of 
$|\F_{AF}\ket$ onto the singlet and the triplet subspace and 
useful identities between the two spin-projected states are obtained. 
Then, in Section \ref{ground} we propose an {\it Ansatz} for the ground 
state wave function at half filling containing the singlet and  
triplet projections of $|\F_{AF}\ket$. We set up the Schr\"odinger 
equation and by exploiting the antiferromagnetic property 
we close the equations and get 3 {\it exact} eigenstates. The 
ground-state uniqueness proved by Lieb\cite{lieb} is used to show that 
the lowest-energy state of our {\it Ansatz} actually corresponds to the 
ground state of the system. Remarkably, the ground state energy is negative for 
any value of the repulsion $U$; qualitatively, we may say that the particles  
manage to  avoid the double occupation very effectively. In Section 
\ref{results} we study the ground state energy $E_{0}$ as a function of $U$ 
and of the volume of the system and we discuss the implications of exchanging the 
thermodynamic limit with the limit of zero temperture. We find that 
$E_{0}$ is a monotonically increasing function of $U$ and $N$ due to 
the existence of non-trivial correlations even for large $N$. The nature 
of these correlations is investigated by computing the expectation 
value of the  repulsion. We show that for any finite $N$ there is a 
critical value of $U$ yielding maximum repulsion. From the exact spin-spin correlation function
we find that the ground state average of the staggered magnetization operator squared is 
{\it exstensive}. Hence, the ground state is antiferromagnetically 
ordered; we underline that this holds not only at strong coupling, but 
for any value of $U$. Finally, a summary of 
the main results and conclusions are drawn in Section \ref{summary}.

\section{Definition of the Model}
\label{model}

Let $\L={\cal A}\cup{\cal B}$ with ${\cal A}\cap{\cal B}=\emptyset$ 
be a collection of sites and 
\begin{equation}
|{\cal A}|=N_{A},\;\;|{\cal B}|=N_{B}\;\Rightarrow |\L|=N_{A}+N_{B}\;.
\end{equation}
Here and in the following we shall denote 
by $|{\cal S}|$ the number of elements in the set ${\cal S}$. We 
consider the Hubbard Hamiltonian 
\begin{equation}
H_{{\mathrm Hubbard}}=H_{0}+W,\;\;\;H_{0}=\sum_{x,y\in\L}\sum_{\s}t_{x,y}c^{\dag}_{x\s}c_{y\s},
\;\;\;W=U\sum_{x\in\L}\hat{n}_{x\ua}\hat{n}_{x\da}
\label{ham}
\end{equation}
where $c_{x\s}$ ($c^{\dag}_{x\s}$) is the annihilation (creation) operator of 
a particle at site $x$ with spin $\s=\ua,\da$ and 
$\hat{n}_{x\s}=c^{\dag}_{x\s}c_{x\s}$ is the corresponding particle 
number operator. The hopping matrix $T$ with elements 
$T_{x,y}=t_{x,y}=t_{y,x}$ is a real-symmetric matrix while $U$ 
is a positive constant. If
\begin{equation}
    t_{x,y}=\left\{\begin{array}{ll}
    \tilde{t} & {\mathrm for}\; x\in{\cal A}\;(x\in{\cal B})\;
    {\mathrm and}\; y\in{\cal B}\;(y\in{\cal A})\\
    0 & {\mathrm otherwise}\end{array}\right.
\end{equation}
the graph $\L$ is said to be complete bipartite and we call the 
Hamiltonian in Eq.(\ref{ham}) the CBG-Hubbard 
Hamiltonian. The model is invariant under an arbitrary 
permutation of the ${\cal A}$-sites and/or of the ${\cal B}$-sites. 
Therefore, the symmetry group includes $S_{N_{A}}\otimes S_{N_{B}}$,  
$S_{N}$ being the set of permutations of $N$ objects. As usual, the full 
group is much bigger: the presence of spin and pseudospin 
symmetries\cite{lieb}\cite{pseudo} leads to an $SO(4)$ internal symmetry 
Group\cite{ya}\cite{yaza} and in the case $N_{A}=N_{B}$ 
there is a $Z_{2}$ 
symmetry because of the ${\cal A} \leftrightarrow {\cal B}$ exchange.

In Fig.(\ref{clust}) we have drawn a few examples of finite-size systems. 
For $N=1$ and $N=2$ the model is equivalent to a one dimensional ring  
of length $L=2,\;4$ respectively. For $N=3$ we have what can be 
understood as a prototype, (1,1) 
{\em nanotube} model, the one of smallest length $L=1$, with periodic 
boundary conditions. For 
general $N$, one can conceive a {\em gedanken} device, like the one 
illustrated  pictorially for $N=6$ in 
Fig.(\ref{clust}.$d$).  (The whole device should 
actually have the topology of a torus, and the two horizontal faces 
should coincide, but this is not shown in the Figure for the sake of 
clarity.) The $N$ vertical lines represent a 
realization of the  ${\cal A}$ sublattice while the  ${\cal B}$ 
sublattice is mimicked by the central object. The radial tracks in the 
Figure represent conducting paths linking 
each ${\cal A}$ site to each ${\cal B}$ site according to the topology 
of our model. The  ${\cal A}$ sites 
are represented by one-dimensional  {\em
electron-in-a-box} systems of length $L$. Each  ${\cal A}$ site has  the one-body energy spectrum 
 $\ve_{{\mathrm free}}\sim n^{2}/m L^{2}$ where $n$ is an integer and 
$m$ is the electron  mass. We assume that $L$ is so small that the 
excited states are at high energy and can be disregarded in the 
low-energy sector; this requires $mL^{2}\ll 1/U$. Here, of course, $U$ 
denotes the Coulomb self-energy of a { \em box} with two electrons. The  ${\cal B}$ sites are hosted by $N$  
quantum dots, that can be 
represented by   $\delta$-function-like attractive 
potential wells of  depth $V$; if  $|V|$ is large and the radius of the dots is $\ll L$, 
these are practically independent of each other.   The one-body energy 
of each ${\cal B}$ site  is  $\ve_{{\mathrm bound}}\sim -mV^{2}$; the 
unbound states can be neglected if  $mV^{2}\gg U$. We are assuming for 
simplicity that the $U$ of the ${\cal B}$ sites is the same as for the  
${\cal A}$ ones. 
  Turning on a constant potential $V_{0}$ on the central object, one can arrange
  that the Hubbard  Hamiltonian on the CBG 
$H_{{\mathrm Hubbard}}$ gives a good description of the  system, with 
filling $\leq 2$ per site.

\begin{figure}[H]
\begin{center}
	\epsfig{figure=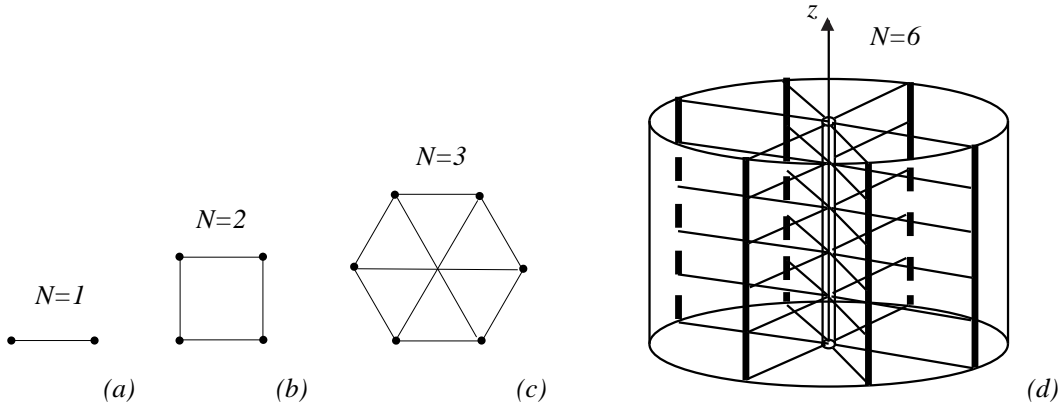,width=14cm}\caption{\footnotesize{
	Physical representation of the CBG for  
	$N=1$ ($a$) and $N=2$ ($b$) which is equivalent to a one dimensional ring  
        of length $L=2,\;4$ respectively. For $N=3$ ($c$) we have the (1,1) 
        nanotube of smallest length and periodic boundary conditions. For 
        $N=6$ ($d$) we present  the {\em gedanken} device described in the 
        text.}}
\label{clust}
\end{center}
\end{figure}
It is convenient to label each site with an integer in such a way that 
$x=1,\ldots,N_{A}$ corresponds to sites in ${\cal A}$ and 
$x=N_{A}+1,\ldots,N_{A}+N_{B}$ corresponds to sites in ${\cal B}$. 
In the special case $N_{A}=N_{B}=N$ the hopping matrix can be written as
\begin{equation}
    T=\tilde{t}\left(\begin{array}{ll} Z & J\\
    J & Z \end{array}\right)
\end{equation}
where $Z$ is the $N\times N$ null  matrix and $J$ is the $N\times N$ matrix whose generic 
element $J_{x,y}=1$. The one-body spectrum has three 
different eigenvalues
\begin{eqnarray}
\ve_{g}&=&-N\tilde{t}\equiv -t \;\;\nonumber \\
\ve_{0}&=&0\;\;\nonumber \\
\ve_{\bar{g}}&=&N\tilde{t}\equiv t
\end{eqnarray}
with degeneracy $d_{g}=1$, $d_{0}=2N-2$ and $d_{\bar{g}}=1$ 
respectively. We use the convention $t>0$ so that $\ve_{g}$ is the 
lowest level and we shall call ${\cal S}_{hf}$ the set of zero-energy 
one-body eigenstates. 

The orthogonal matrix $O$ that diagonalizes $T$ 
can be written in the form 
\begin{equation}
O=\left(\begin{array}{ll}
\begin{array}{c} 
    \frac{1}{\sqrt{2N}}\;\frac{1}{\sqrt{2N}}\ldots\\
    R \\ 
    Z_{r} \\
    \frac{1}{\sqrt{2N}}\;\frac{1}{\sqrt{2N}}\ldots
\end{array} &
\begin{array}{c} 
    \frac{-1}{\sqrt{2N}}\;\frac{-1}{\sqrt{2N}}\ldots\\
    Z_{r} \\ 
    R \\
    \frac{1}{\sqrt{2N}}\;\frac{1}{\sqrt{2N}}\ldots
\end{array} 
\end{array}\right)
\label{ort}
\end{equation}
where $Z_{r}$ is the $(N-1)\times N$ rectangular null matrix,
while $R$ is an $(N-1)\times N$ rectangular matrix whose rows 
are orthonormal vectors which are orthogonal to the $N-$dimensional vector $(1,1,\ldots,1,1)$. 
The zero-energy one-body orbitals 
can be visualized by a simple argument. Consider 
any pair $x,\;y$, with $x \neq y,$ of sites belonging to the same sublattice, say ${\cal A}$, 
and a wavefunction $\psi_{x,y}$  taking the values 1 and -1 on the pair
and 0 elsewhere in ${\cal A}$ and in  ${\cal B}$. It is evident 
that $\psi_{x,y}$ belongs to ${\cal S}_{hf}$. Operating on  $\psi_{x,y}$ 
by the permutations of $S_{N}$ we can generate a (non-orthogonal) 
basis of $N-1$ eigenfunctions\footnote{
There is a simple trick to see that the number of independent 
wavefunctions obtained in this way  is $N-1$.  Let $P \in S_{N}$ be 
a permutation such that $P(x)=x$ and $P(y) \neq x$. We get the full 
basis since $P(y)$ takes $N-1$ values with independent 
$\psi_{x,P(y)}$.} vanishing in ${\cal B}$; further, 
by means of the $Z_{2}$ symmetry,  we obtain the remaining orbitals 
of ${\cal S}_{hf}$, which vanish on ${\cal A}$. This exercise shows 
that the group considered above justifies the ($2N-2$)-fold degeneracy 
of the one-body spectrum.

Having obtained the one-body basis, one can form a suitable $R$ 
matrix by orthogonalization. Let us define the transformed operators:
\begin{eqnarray}
g\equiv\sum_{x=1}^{2N}O_{1,x}c_{x}=\frac{1}{\sqrt{2N}}\sum_{x=1}^{N}(c_{x}-c_{x+N})
\nonumber \\
a_{i}\equiv\sum_{x=1}^{2N}O_{i+1,x}c_{x}=\sum_{x=1}^{N}R_{i,x}c_{x},\;\;
i=1,\ldots,N-1
\nonumber \\
b_{i}\equiv\sum_{x=1}^{2N}O_{i+N,x}c_{x}=\sum_{x=1}^{N}R_{i,x}c_{x+N},\;\;
i=1,\ldots,N-1
\nonumber \\
\bar{g}\equiv\sum_{x=1}^{2N}O_{2N,x}c_{x}=\frac{1}{\sqrt{2N}}
\sum_{x=1}^{N}(c_{x}+c_{x+N}).
\label{traop}
\end{eqnarray}
The inverse transformation reads
\begin{eqnarray}
    c_{x}=\frac{1}{\sqrt{2N}}(g+\bar{g})+\sum_{i}R_{i,x}a_{i},
   \;\; x \in {\cal A}\nonumber\\
    c_{x}=\frac{1}{\sqrt{2N}}(-g+\bar{g})+\sum_{i}R_{i,x-N}b_{i},
   \;\; x \in {\cal B}
\label{inve}
\end{eqnarray}   
and the kinetic term $H_{0}$ becomes
\begin{equation}
H_{0}=-t\sum_{\s}(g^{\dag}_{\s}g_{\s}-\bar{g}^{\dag}_{\s}\bar{g}_{\s})\;.
\label{hzerodia}
\end{equation}
Hence, if we do not rescale the hopping constant, 
the average kinetic energy remains an extensive quantity proportional 
to $N\tilde{t}$: the kinetic energy of the two particles in the lowest 
level coincides with the kinetic energy of the whole system. On the 
other hand, if $\tilde{t}\sim 1/N$ the two energy gaps $\ve_{0}-\ve_{g}$ and 
$\ve_{\bar{g}}-\ve_{0}$ remain constant as $N$ increases. As we shall 
show in the next Section, the model can be exactly solved in the 
thermodynamic limit whatever is the dependence of $\ve_{g}$ and 
$\ve_{\bar{g}}$ on $|\L|$. 

\section{Thermodynamic Limit}
\label{thermolimi}

The Hubbard Hamiltonian on the CBG (as on the complete 
graph) belongs to a class of Hubbard-like models that can be exactly 
solved in the thermodynamic limit. Here we generalize the results by 
van Dongen and Vollhardt\cite{vDV} to the case of a non-extensive number of 
non-vanishing one-body eigenenergies and in the presence of a local 
external magnetic field coupled to the spin of the particles. Even if 
the reasoning is similar to that of Ref.\cite{vDV}, a detailed presentation of the main results
is needed to clarify more subtle 
points and make our  work self-contained. Furthermore, some of the 
results are absent in the original paper and we believe that their 
 derivation will facilitate the reader in the comprehension of 
what follows. 

Let us consider the Hamiltonian 
\begin{equation}
H=H_{{\mathrm Hubbard}}-\m\sum_{x\in\L}\sum_{\s}\hat{n}_{x\s}-
\sum_{x\in\L}h_{x}(\hat{n}_{x\ua}-\hat{n}_{x\da})\equiv
H_{{\mathrm Hubbard}}+H_{\m}+H_{h}
\label{hamgen}
\end{equation}
where $H_{{\mathrm Hubbard}}=H_{0}+W$ is defined in Eq.(\ref{ham}), 
$\m$ is the chemical potential and $H_{h}$ represents the coupling 
with an external magnetic field. Let us assume that only a finite 
number f of eigenvalues of the hopping matrix $T$ are 
non-vanishing, $\ve_{i}\neq 0$ for $i=1,\ldots,$f, while the 
remaining $|\L|-$f are identically zero. Denoting with $\f_{i}$ and 
$\f^{\dag}_{i}$ the annihilation and creation operators that 
diagonalize $H_{0}$ we have
\begin{equation}
H_{0}=\sum_{i=1}^{{\mathrm f}}\sum_{\s}\ve_{i}\f^{\dag}_{i\s}\f_{i\s}
\end{equation}
and the CBG   corresponds to the case f=2, 
see Eq.(\ref{hzerodia}). The Gran-Partition Function 
${\cal Z}_{\b}[T,U,\{h_{x}\}]$ can always be written as
\begin{equation}
{\cal Z}_{\b}[T,U,\{h_{x}\}]={\cal Z}_{\b}[T,0,\{h_{x}=0\}]
e^{{\cal W}_{\b}[T,U,\{h_{x}\}]},\;\;\;\;
{\cal W}_{\b}[T,U,\{h_{x}\}]=
\bra e^{-\int_{0}^{\b}H_{I}(\t)d\t}\ket_{0}^{c}
\label{therm1}
\end{equation}
where $\b$ is the inverse Temperature, $H_{I}(\t)=W(\t)+H_{h}(\t)$ is the 
interaction Hamiltonian in the imaginary-time Heisenberg 
representation and $\bra \ldots\ket_{0}^{c}$ is the thermal average in 
terms of ${\cal Z}_{\b}[T,0,\{h_{x}=0\}]\equiv {\cal Z}_{\b}^{0}[T]$ 
where, by the linked-cluster theorem, one has to retain only those 
contributions represented by connected diagrams. As far as f remains 
fixed, we can safely substitute 
${\cal W}_{\b}[T,U,\{h_{x}\}]$ with ${\cal W}_{\b}[T=0,U,\{h_{x}\}]$  
in Eq.(\ref{therm1}). Indeed,  
\begin{equation}
{\cal W}_{\b}[T,U,\{h_{x}\}]={\cal W}_{\b}[T=0,U,\{h_{x}\}]+
O({\mathrm f}/|\L|)\;
\label{simpl}
\end{equation}
as long as the unperturbed one-body Green's functions $G_{i\s}(\t)=
\bra{\cal T}\f_{i\s}(\t)\f^{\dag}_{i\s}(0)\ket_{0}$, $i=1,\ldots,$f, 
do not diverge in the thermodynamic limit; here ${\cal T}$ is the Wick 
time-ordering operator and $\bra\ldots\ket_{0}$ is the thermal average 
with $U=0$ and $\{h_{x}=0\}$. From the explicit expression of  
$G_{i\s}(\t)$, 
\begin{equation}
G_{i\s}(\t)=\left\{\begin{array}{ll}
e^{-(\ve_{i}-\m)\t}(1-f_{i}) & \t>0 \\
-e^{-(\ve_{i}-\m)\t}f_{i} & \t\leq 0
\end{array}\right.\;,
\end{equation}
with $f_{i}=[1+e^{\b(\ve_{i}-\m)}]^{-1}$ the Fermi distribution 
function, it follows that $G_{i\s}(\t)$ converges to a finite 
value whenever $\lim_{|\L|\ra\inf}\ve_{i}$ is well defined. One can 
use the result in Eq.(\ref{simpl}) to write the exact Gran-Partition 
Function in the thermodynamic limit
\begin{eqnarray}
{\cal Z}_{\b}[T,U,\{h_{x}\}]&=&\frac{{\cal Z}_{\b}^{0}[T]}
{{\cal Z}_{\b}^{0}[T=0]}{\cal Z}_{\b}[T=0,U,\{h_{x}\}]=
\nonumber \\ &=&
\prod_{i=1}^{{\mathrm f}}\left[
\frac{1+2ze^{-\b\ve_{i}}+z^{2}e^{-2\b\ve_{i}}}{(1+z)^{2}}\right]
\prod_{x\in\L}\left[
1+2z\cosh \b h_{x}+z^{2}e^{-\b U}\right]\;,
\end{eqnarray}
where $z=e^{\b\m}$ is the fugacity, and the thermodynamic potential 
$\W_{\b}[T,U,\{h_{x}\}]=-\b^{-1}\ln {\cal 
Z}_{\b}[T,U,\{h_{x}\}]$
\begin{equation}
\W_{\b}[T,U,\{h_{x}\}]=\frac{1}{\b}\left\{
2{\mathrm f}\ln(1+z)-\sum_{i=1}^{{\mathrm f}}
\ln[1+2ze^{-\b\ve_{i}}+z^{2}e^{-2\b\ve_{i}}]-
\sum_{x\in\L}\ln\left[1+2z\cosh \b h_{x}+z^{2}e^{-\b U}\right]\right\}.
\;\;
\label{therm2}
\end{equation}
Eq.(\ref{therm2}) reduces to the result of van Dongen and Vollhardt 
in the case f=1, $h_{x}=0$ $\forall x\in\L$ and 
$\ve_{1}\propto -|\L|<0$. 
All the thermodynamic quantities can be derived from 
Eq.(\ref{therm2}) and we defer the interested reader to the original 
paper by  van Dongen and Vollhardt for further details. Here we 
calculate the magnetization at site $x$, $m_{x}=-\frac{1}{2}(\de\W_{\b}/\de h_{x})=
\frac{1}{2}\bra \hat{n}_{x\ua}-\hat{n}_{x\da}\ket\equiv \bra 
S^{z}_{x}\ket$ and the connected spin-spin  
correlation function $\bra S^{z}_{x}S^{z}_{y}\ket^{c}\equiv 
\bra S^{z}_{x}S^{z}_{y}\ket-\bra S^{z}_{x}\ket\bra S^{z}_{y}\ket =
-(1/4\b)(\de^{2}\W_{\b}/\de h_{x}\de h_{y} )$, where 
$\bra\ldots\ket$ means the average in the gran-canonical ensemble. Due 
to the exact decoupling of the kinetic energy, we expect from the 
begining a paramagnetic behaviour. We have
\begin{equation}
m_{x}=-\frac{1}{2}\frac{\de\W_{\b}}{\de h_{x}}=\frac{z\sinh \b 
h_{x}}{1+2z\cosh \b h_{x}+z^{2}e^{-\b U}}
\label{emme}
\end{equation}
where $z$ can be expressed in terms of the number density by means of 
the relation
\begin{equation}
n=-\frac{1}{|\L|}\frac{\de\W_{\b}}{\de\m}=
\frac{2z}{|\L|}\sum_{x\in\L}\frac{\cosh \b h_{x}+ze^{-\b U}}
{1+2z\cosh \b h_{x}+z^{2}e^{-\b U}}\;.
\label{zeta}
\end{equation}
Since $z$ is finite for any non-zero Temperature, $m_{x}\ra 0$ when 
$h_{x}\ra 0$, {\it i.e.} 
\begin{equation}
\lim_{\b\ra \inf}\lim_{h_{x}\ra 0}\lim_{|\L|\ra\inf}
m_{x}=0\;.
\label{limi}
\end{equation}
On the other hand, $m_{x}\neq 0$ if we exchange the last two limits 
in Eq.(\ref{limi}). Indeed, taking $h_{x}=h$=const$>0$ for the sake of 
clarity, Eq.(\ref{zeta}) can be exactly solved for $z$:
\begin{equation}
z=\frac{\sqrt{(n-1)^{2}\cosh^{2} \b h+n(2-n)e^{-\b U}}-(1-n)\cosh\b h}
{(2-n)e^{-\b U}}\ra_{\b\ra\inf}
\left\{\begin{array}{ll}
e^{\b(U+h)}(n-1)/(2-n) & n>1 \\
e^{\b U/2}[n/(2-n)]^{1/2} & n=1 \\
e^{-\b h}n/(1-n) & n<1
\end{array}\right.\;.
\end{equation}
Substitution of these asymptotic behaviours in Eq.(\ref{emme}) yields
\begin{equation}
\lim_{h_{x}\ra 0^{+}}\lim_{\b\ra \inf}\lim_{|\L|\ra\inf}
m_{x}= \left\{\begin{array}{ll}
(2-n)/2 & n>1 \\
1/2 & n=1 \\
n/2 & n<1 \end{array}\;.
\right.
\label{magn}
\end{equation}
The case $h<0$ is similar and one can show that 
Eq.(\ref{magn}) remains true in the limit 
$h_{x}\ra 0^{-}$  if $m_{x}\ra -m_{x}$. 
Therefore, the magnetization does not depend on $U$ and the 
zero-Temperature limit of the model is the same of the 
paramagnetic Hamiltonian $H_{{\mathrm para}}=H_{\m}+H_{h}$, where 
$H_{\m}$ and $H_{h}$ are defined in Eq.(\ref{hamgen}). From 
Eq.(\ref{emme}) we also conclude that 
\begin{equation}
\bra S^{z}_{x}S^{z}_{y}\ket^{c}\equiv 
-\frac{1}{4\b}\left(\frac{\de^{2}\W_{\b}}{\de h_{x}\de h_{y}}
\right) =0\;\;\Rightarrow 
\bra S^{z}_{x}S^{z}_{y}\ket=\bra S^{z}_{x}\ket\bra S^{z}_{y}\ket=
m_{x}m_{y},
\end{equation}
a result which is independent of the local external field configuration: in the 
thermodinamic limit two localized spins do not interact and the 
spin-spin correlation length is zero. Denoting with $G_{{\mathrm 
spin}}(x,y;\b,h,|\L|)$ 
the thermal average of $S^{z}_{x}S^{z}_{y}$ at inverse Temperature $\b$, 
external field $h$ and size of the system $|\L|$, from Eq.(\ref{limi}) 
we get  
\begin{equation}
\lim_{\b\ra \inf}\lim_{h\ra 0}\lim_{|\L|\ra\inf}
\bra S^{z}_{x}S^{z}_{y}\ket=
\lim_{\b\ra \inf}\lim_{h\ra 0}\lim_{|\L|\ra\inf}
G_{{\mathrm spin}}(x,y;\b,h,|\L|)=0\;,
\label{corr}
\end{equation}
that is, the spin-spin correlation function vanishes if we first take 
the limit $h\ra 0$ and then the limit of $\b\ra\inf$. 
We emphasize that the above results are correct only if we 
first take the limit $|\L|\ra \inf$. The thermodynamic 
limit makes the model trivial and different graph structures, like the 
complete graph and the complete bipartite graph, all yield the same 
thermodynamic behaviour. On the contrary, a much more difficult task 
consists in finding {\it exact} results in finite-size systems. To face  
such problems is not only a mathematical exercise. After that in Section 
\ref{ground} the {\it exact} ground state of the 
half-filled Hubbard model on the CBG is explicitly 
written down for arbitrary value of $U$ and $|\L|$ and in the absence 
of an external field $h$,  we shall see in Section \ref{results} that 
the thermodynamic limit, $|\L|\ra\inf$, and the limits of zero Temperature 
and external field, $\b\ra\inf,\;h\ra 0$, do not commute. 

\section{The $W=0$ States}
\label{w=0states}

In this Section we introduce the essential tools to deal with the 
antiferromagnetic ground-state solution. Let us consider 
the one-body eigenstate $a^{\dag}_{i}|0\ket$ with vanishing amplitudes 
on the ${\cal B}$ sublattice, that is $\bra 0|c_{x}a^{\dag}_{i}|0\ket=0$ if 
$x\in {\cal B}$. Similarly, $b^{\dag}_{i}|0\ket$ has vanishing amplitude on the 
${\cal A}$ sublattice and therefore the $(2N-2)$-body state 
\begin{equation}
|\F_{AF}^{(\s)}\ket=a^{\dag}_{1\s}\ldots a^{\dag}_{N-1\s}
b^{\dag}_{1\bar{\s}}\ldots b^{\dag}_{N-1\bar{\s}}|0\ket,\;\;\;\;
\bar{\s}=-\s
\label{detaf}
\end{equation}
is an eigenstate of $H_{0}$ and of $W$ with vanishing eigenvalue. In 
the following we shall use the wording {\it $W=0$ state} to denote any 
eigenstate of $H_{{\mathrm Hubbard}}$ in the kernel of $W$. It is worth 
to observe that by mapping the ${\cal A}$-sites 
onto the ${\cal B}$-sites and {\it viceversa}, $|\F_{AF}\ket$ retains 
its form except for a spin-flip; we  call this property the {\it 
antiferromagnetic property} for obvious reasons.

In the non-interacting ($U=0$) half filled system, the structure of 
the ground state is trivial: two particles 
of opposite spin sit in the lowest energy level $g$, while $|{\cal 
S}_{hf}|=(2N-2)$ 
particles lie on the shell ${\cal S}_{hf}$ of zero energy. 
In the spin $S^{z}=0$ subspace this ground state is 
$\left(\begin{array}{c} 2N-2\\ N-1\end{array}\right)^{2}$ times 
degenerate. To first order in $W$, the degeneracy is only partially 
removed\footnote{At weak coupling it makes sense to speak 
about particles in filled shells, which behave much as core electrons 
in atomic physics, and particles in partially filled, or valence, shells.
In first order, we must diagonalize the perturbation $W$ in the 
degenerate subspace, dropping the filled shells. Since $W$ is a 
positive semi-definite operator, with vanishing eigenvalue, 0 is 
necessarily the lowest one. Typically, the first-order solution is a 
crude approximation to the exact one, but below, when discussing the 
exact results, we shall demonstrate that many of the qualitative features 
hold for all $U$.}. Indeed, if $P_{S,M_{S}}$ is the projection operator onto the 
subspace of total spin $S$ with $z$-component $M_{S}$, the structure 
of the determinantal state in Eq.(\ref{detaf}) yields
\begin{equation}
P_{S,0}|\F_{AF}^{(\s)}\ket\equiv|\F_{AF}^{S,0}\ket
\neq 0,\;\;\forall S=0,\ldots,N-1\;;
\end{equation}
therefore, the states $\{g^{\dag}_{\ua}g^{\dag}_{\da}|\F_{AF}^{S,0}\ket,
\;S=0,\ldots,N-1\}$ belong to the ground state multiplet in first order 
perturbation theory (being $W=0$ states). Since 
the Lieb's theorem\cite{lieb} ensures that the interacting ground state 
$|\Q_{0}(N,U)\ket$ 
is a singlet, only $g^{\dag}_{\ua}g^{\dag}_{\da}|\F_{AF}^{0,0}\ket$ 
can have a non-vanishing overlap with $|\Q_{0}(N,U)\ket$ in the limit 
$U\ra 0^{+}$. In Section \ref{ground} we shall prove that 
$g^{\dag}_{\ua}g^{\dag}_{\da}|\F_{AF}^{0,0}\ket$ is the 
ground state for $U=0^{+}$, that is $\lim_{U\ra 0}\bra\Q_{0}(N,U)|
g^{\dag}_{\ua}g^{\dag}_{\da}|\F_{AF}^{0,0}\ket=1$.  In this 
model, the first-order solution has a peculiar significance. Usually, 
the exact ground state has no overlap with  the non-interacting one, 
in the thermodynamic limit: this is because the $U=0$ ground states 
define a {\em proper} subspace of the full Hilbert space. Below, we 
prove that $\bra\Q_{0}(N,U)|g^{\dag}_{\ua}g^{\dag}_{\da}|\F_{AF}^{0,0}\ket 
\neq 0 \; \forall U$; that is, the lowest approximation keeps a finite weight. 

At this stage, 
we note  the analogy of the above results with those relevant to the 
standard Hubbard model defined on a $N\times N$ square lattice with periodic 
boundary conditions (hopping matrix elements only between nearest neighbor 
sites). The determinantal state $|\F_{AF}^{(\s)}\ket$ resembles the $W=0$ 
state that  we obtained for even $N$  in Ref.\cite{ssc2001}\cite{jop2001}.  
The degeneracy of the zero-energy one-body eigenspace was $2N-2$ and it was 
shown\cite{jop2002} that $N-1$ zero-energy eigenfunctions have vanishing 
amplitudes on a sublattice while the remaining $N-1$ vanish on the other.

As we shall see in the next Section, we need the explicit expression for the 
singlet $|\F_{AF}^{0,0}\ket$ and the triplet $|\F_{AF}^{1,m}\ket$, 
$m=0,\pm 1$, to calculate the ground state of 
the CBG-Hubbard model. Therefore, we now briefly review how to get the projection  
of the determinantal state $|\F_{AF}^{(\s)}\ket$ onto the singlet and 
the triplet spin subspaces. 

To obtain the singlet $|\F_{AF}^{0,0}\ket$ one has to 
antisymmetrize each product $a^{\dag}_{i\ua}b^{\dag}_{j\da}$, getting a 
two-body spin singlet operator, and subsequently antisymmetrize with respect 
to the $N-1$ indices of the $b^{\dag}$'s; one sees easily that this 
entails the antisymmetrization of the $a^{\dag}$'s. 
Hence 
\begin{equation}
|\F_{AF}^{0,0}\ket=\sum_{P}(-)^{P}\prod_{i=1}^{N-1}\s^{\dag}_{i,P(i)}
|0\ket
\label{singaf}
\end{equation}
where $\s^{\dag}_{i,j}$ creates a two-body singlet state:
\begin{equation}
\s^{\dag}_{i,j}=\frac{1}{\sqrt{2}}(a^{\dag}_{i\ua}b^{\dag}_{j\da}-
a^{\dag}_{i\da}b^{\dag}_{j\ua})\;,
\end{equation}
and $P$ is a permutation of the indices $1,\ldots,N-1$. Since $W$ 
commutes with the total spin operators and the determinant $|\F_{AF}^{(\s)}\ket$ 
is a $W=0$ state, $W|\F_{AF}^{0,0}\ket=0$ as already noted. Applying $W$ 
on the state in Eq.(\ref{singaf}), 
one can verify this property by direct inspection.  

In a similar way one obtains the triplet projection. We define 
$\t^{(0)^{\dag}}_{i,j}$ as the two-body 
triplet creation operator with vanishing $z$-component $m =0$ in the orbitals 
$a_{i}$ and $b_{j}$:
\begin{equation}
\t^{(0)^{\dag}}_{i,j}=\frac{1}{\sqrt{2}}(a^{\dag}_{i\ua}b^{\dag}_{j\da}+
a^{\dag}_{i\da}b^{\dag}_{j\ua})\;.
\end{equation}
Then, one has 
\begin{equation}
|\F_{AF}^{1,0}\ket=\sum_{P}(-)^{P}\sum_{j=1}^{N-1}
\t^{(0)^{\dag}}_{j,P(j)}
\prod_{i\neq j}\s^{\dag}_{i,P(i)}
|0\ket\;.
\label{trip0af}
\end{equation}
The $m=\pm 1$ components of the triplet in Eq.(\ref{trip0af}) can 
be otained by means of the total raising and lowering spin 
operators. Let us introduce the following notations 
\begin{eqnarray}
S^{+a}_{i}=a^{\dag}_{i\ua}a_{i\da},\;\;\;S^{-a}_{i}=[S^{+a}_{i}]^{\dag},
\;\;\;S^{+a}=\sum_{i=1}^{N-1}S^{+a}_{i},\;\;\;S^{-a}=[S^{+a}]^{\dag}
\\ 
S^{+b}_{i}=b^{\dag}_{i\ua}b_{i\da},\;\;\;S^{-b}_{i}=[S^{+b}_{i}]^{\dag},
\;\;\;S^{+b}=\sum_{i=1}^{N-1}S^{+b}_{i},\;\;\;S^{-b}=[S^{+b}]^{\dag}\;
 \\
\hat{n}^{a}_{i\s}=a^{\dag}_{i\s}a_{i\s},\;\;\;
S^{za}_{i}=\frac{1}{2}(\hat{n}^{a}_{i\ua}-\hat{n}^{a}_{i\da}),\;\;\;
\hat{n}^{a}_{\s}=\sum_{i=1}^{N-1}\hat{n}^{a}_{i\s},\;\;\;
\hat{n}^{a}=\hat{n}^{a}_{\ua}+\hat{n}^{a}_{\da},\;\;\;
S^{za}=\frac{1}{2}(\hat{n}^{a}_{\ua}-\hat{n}^{a}_{\da})\\
\hat{n}^{b}_{i\s}=b^{\dag}_{i\s}b_{i\s},\;\;\;
S^{zb}_{i}=\frac{1}{2}(\hat{n}^{b}_{i\ua}-\hat{n}^{b}_{i\da}),\;\;\;
\hat{n}^{b}_{\s}=\sum_{i=1}^{N-1}\hat{n}^{b}_{i\s},\;\;\;
\hat{n}^{b}=\hat{n}^{a}_{\ua}+\hat{n}^{b}_{\da},\;\;\;
S^{zb}=\frac{1}{2}(\hat{n}^{b}_{\ua}-\hat{n}^{b}_{\da})\\\;.
\end{eqnarray}
The states $|\F_{AF}^{1,\pm 1}\ket$ can be written as 
\begin{eqnarray}
|\F_{AF}^{1,1}\ket=\frac{1}{\sqrt{2}}(S^{+a}+S^{+b})|\F_{AF}^{1,0}\ket=
\sum_{P}(-)^{P}\sum_{j=1}^{N-1}
\t^{(+1)^{\dag}}_{j,P(j)}
\prod_{i\neq j}\s^{\dag}_{i,P(i)}
|0\ket 
\label{trip1af}\\
|\F_{AF}^{1,-1}\ket=\frac{1}{\sqrt{2}}(S^{-a}+S^{-b})|\F_{AF}^{1,0}\ket=
\sum_{P}(-)^{P}\sum_{j=1}^{N-1}
\t^{(-1)^{\dag}}_{j,P(j)}
\prod_{i\neq j}\s^{\dag}_{i,P(i)}
|0\ket
\label{trip-1af}
\end{eqnarray}
with
\begin{equation}
\t^{(+1)^{\dag}}_{i,j}=a^{\dag}_{i\ua}b^{\dag}_{j\ua},\;\;\;\;\;
\t^{(-1)^{\dag}}_{i,j}=a^{\dag}_{i\da}b^{\dag}_{j\da}\;.
\end{equation}
Equivalently, the triplet state $|\F_{AF}^{1,m}\ket$, $m=0,\pm 1$, can also be 
expressed in terms 
of the singlet $|\F_{AF}^{0,0}\ket$. It is a simple exercise to 
prove the following identities
\begin{eqnarray}
\sum_{i=1}^{N-1}(S^{+a}_{i}-S^{+b}_{i})|\F_{AF}^{0,0}\ket=
(S^{+a}-S^{+b})|\F_{AF}^{0,0}\ket=
-\sqrt{2}|\F_{AF}^{1,1}\ket
\label{id1}\\
\sum_{i=1}^{N-1}(\hat{n}^{a}_{i\da}-\hat{n}^{b}_{i\da})|\F_{AF}^{0,0}\ket=
\sum_{i=1}^{N-1}(-\hat{n}^{a}_{i\ua}+\hat{n}^{b}_{i\ua})|\F_{AF}^{0,0}\ket=
(-S^{za}+S^{zb})|\F_{AF}^{0,0}\ket=
-|\F_{AF}^{1,0}\ket
\label{id2}\\
\sum_{i=1}^{N-1}(-S^{-a}_{i}+S^{-b}_{i})|\F_{AF}^{0,0}\ket=
(-S^{-a}+S^{-b})|\F_{AF}^{0,0}\ket=
-\sqrt{2}|\F_{AF}^{1,-1}\ket\;.
\label{id3}
\end{eqnarray}
As for the singlet, one can verify that $W|\F_{AF}^{1,m}\ket=0$, 
$m=0,\pm 1$, using the definitions in  
Eqs.(\ref{trip0af}-\ref{trip1af}-\ref{trip-1af}).

\section{The Ground State at Half Filling}
\label{ground}

We are now ready to calculate the ground state $|\Q_{0}(N,U)\ket$ of 
the half filled CBG-Hubbard model 
described in Section \ref{model}. As already observed, 
$g^{\dag}_{\ua}g^{\dag}_{\da}|\F_{AF}^{0,0}\ket$ is a good candidate 
for the non-interacting ground state; its quantum numbers are the same 
as those
of $|\Q_{0}(N,U)\ket$ and moreover it belongs to the first-order
ground state multiplet. Using the short-hand notations
\begin{equation}
|g^{0}\ket\equiv g^{\dag}_{\ua}g^{\dag}_{\da}|0\ket,\;\;\;\;\;
|\bar{g}^{0}\ket\equiv \bar{g}^{\dag}_{\ua}\bar{g}^{\dag}_{\da}|0\ket,\;\;\;\;\;
|[g\bar{g}]^{0,0}\ket\equiv\frac{1}{\sqrt{2}}(g^{\dag}_{\ua}\bar{g}^{\dag}_{\da}-
g^{\dag}_{\da}\bar{g}^{\dag}_{\ua})|0\ket
\end{equation}
for the three different two-body singlets that one gets from the lowest and 
the highest energy orbitals $g$ and $\bar{g}$ and
\begin{equation}
|[g\bar{g}]^{1,1}\ket\equiv g^{\dag}_{\ua}\bar{g}^{\dag}_{\ua}|0\ket,
\;\;\;\;\;
|[g\bar{g}]^{1,0}\ket\equiv 
\frac{1}{\sqrt{2}}(g^{\dag}_{\ua}\bar{g}^{\dag}_{\da}+
g^{\dag}_{\da}\bar{g}^{\dag}_{\ua})|0\ket,\;\;\;\;\;
|[g\bar{g}]^{1,-1}\ket\equiv g^{\dag}_{\da}\bar{g}^{\dag}_{\da}|0\ket
\end{equation}
for the triplet, we propose the following {\it Ansatz} 
for the interacting ground state $|\Q_{0}(N,U)\ket$:
\begin{equation}
|\Q_{0}(N,U)\ket=\left[\g_{g}|g^{0}\ket+\g_{\bar{g}}|\bar{g}^{0}\ket\right]
\otimes|\F_{AF}^{0,0}\ket+
\g_{0}\sum_{m=-1}^{1}(-)^{m}|[g\bar{g}]^{1,m}\ket\otimes|\F_{AF}^{1,-m}\ket\;,
\label{ansatz}
\end{equation}
where the $\gamma$'s are $c$-numbers.
It is worth to note that in $|\Q_{0}(N,U)\ket$ the number of particles 
in the shell ${\cal S}_{hf}$ is a constant given by $2N-2$. 
{\it A priori}, there is no reason for this choice   
since the total number operator $\hat{n}^{a}+
\hat{n}^{b}$ of particles in the shell ${\cal S}_{hf}$ 
does not commute with the Hamiltonian. Nevertheless, we 
shall see that the scatterings which do not preserve this number cancel 
out provided in ${\cal S}_{hf}$ the $(2N-2)$-body state is a $W=0$ state. 
We shall refer to this very remarkable property 
as to the {\it Shell Population Rigidity}. We emphasize that even constraining the 
particle-number in ${\cal S}_{hf}$ to be $2N-2$ with vanishing total spin 
$z$-component, there are 
$\left(\begin{array}{c} 2N-2\\ N-1\end{array}\right)^{2}$  
configurations that can contribute to the ground state expansion in 
the interacting case, while the {\it Ansatz} (\ref{ansatz}) contains only  
$|\F_{AF}^{0,0}\ket$ and $|\F_{AF}^{1,m}\ket$, $m=0,\pm 1$. 
The reason why the $W=0$ states $|\F_{AF}^{S,M_{S}}\ket$ 
with $S>1$ do not enter into the ground state expansion (\ref{ansatz}) 
comes from the Lieb's theorem\cite{lieb}: the ground state 
must be a singlet and with only two particles outside ${\cal S}_{hf}$ 
(in the $g$ and/or $\bar{g}$ states) the angular momentum composition 
law  forbids $W=0$ states with $S>1$. However, we observe that the 
state $|[g\bar{g}]^{0,0}\ket\otimes |\F_{AF}^{0,0}\ket$ is a singlet 
and $|\F_{AF}^{0,0}\ket$ is a $W=0$ state. It has the right quantum 
numbers and, in principle, it could have a non-zero overlap with $|\Q_{0}(N,U)\ket$. 
Nevertheless, the matrix elements of $W$ between this state and the ones in the {\it 
Ansatz} are zero. This is the reason why we have dropped it in the expansion 
(\ref{ansatz}). 

The three states in Eq.(\ref{ansatz}) are eigenstates of the 
kinetic-energy operator $H_{0}$:
\begin{eqnarray}
&H_{0}&|g^{0}\ket\otimes|\F_{AF}^{0,0}\ket=-2t|g^{0}\ket\otimes|\F_{AF}^{0,0}\ket
\nonumber \\ \nonumber\\
&H_{0}&|\bar{g}^{0}\ket\otimes|\F_{AF}^{0,0}\ket=2t
|\bar{g}^{0}\ket\otimes|\F_{AF}^{0,0}\ket\nonumber\\ 
&H_{0}&\sum_{m=-1}^{1}(-)^{m}|[g\bar{g}]^{1,m}\ket\otimes|\F_{AF}^{1,-m}\ket=0\;.
\label{kinact}
\end{eqnarray}

Expanding $c_{x \da}c_{x \ua}$ occurring in $W$ in $a,\;b,\;g,\; \bar{g}$ 
operators of Eq.(\ref{traop}), we may take advantage 
from the fact that 
$|\F_{AF}^{0,0}\ket$ and $|\F_{AF}^{1,m}\ket$, $m=0,\pm 1$, are $W=0$ 
states. Indeed, since one cannot annihilate $g$ or $\bar{g}$ over them, taking 
for instance $x \in {\cal A}$,
\begin{equation}
0=c_{x \da}c_{x \ua}|\F_{AF}^{S,M_{S}}\ket=   
\sum_{i,j}R_{i,x}R_{j,x}a_{i \da}a_{j \ua}
|\F_{AF}^{S,M_{S}}\ket\;;
\label{no2a}
\end{equation}
then, multiplying by $g_{\ua}^{\dagger}g_{\da}^{\dagger}$, and the 
like, one obtains that the contribution proportional to    
\begin{equation}
\sum_{i,j}R_{i,x}R_{j,x}a_{i \da}a_{j \ua}|\Q_{0}(N,U)\ket
\label{mancava}
\end{equation}
yield nothing. Of course, if $x \in {\cal B}$, by the same reasoning 
we get 
\begin{equation}
\sum_{i,j}R_{i,x}R_{j,x}b_{i \da}b_{j \ua}|\Q_{0}(N,U)\ket=0.
\label{no2b}
\end{equation}
Hence, we can write 
\begin{equation}
    W=W_{{\cal A}}+W_{{\cal B}}
\end{equation}
with
\begin{equation}
W_{{\cal A}}:=U\sum_{x=1}^{N}c^{\dag}_{x\ua}c^{\dag}_{x\da}\left[
\frac{1}{2N}(g_{\da}+\bar{g}_{\da})(g_{\ua}+\bar{g}_{\ua})+
\frac{1}{\sqrt{2N}}\sum_{i=1}^{N-1}R_{i,x}[(g_{\da}+\bar{g}_{\da})a_{i\ua}+
a_{i\da}(g_{\ua}+\bar{g}_{\ua})]
\right]
\end{equation}
and
\begin{equation}
W_{{\cal B}}:=U\sum_{x=N+1}^{2N}c^{\dag}_{x\ua}c^{\dag}_{x\da}\left[
\frac{1}{2N}(-g_{\da}+\bar{g}_{\da})(-g_{\ua}+\bar{g}_{\ua})+
\frac{1}{\sqrt{2N}}\sum_{i=1}^{N-1}R_{i,x-N}[(-g_{\da}+\bar{g}_{\da})b_{i\ua}+
b_{i\da}(-g_{\ua}+\bar{g}_{\ua})
\right]\;,
\end{equation}
where $:=$ means that the two sides are equivalent
when acting on $|\Q_{0}(N,U)\ket$.
Next, we transform the remaining two $c^{\dag}$ operators, using 
Eq.(\ref{inve}). We note that the terms containing the sequence 
$g^{\dagger}g^{\dagger}g \;a$ and similar terms with $b$ instead of $a$ 
and/or some  $g$ replaced by  $\bar{g}$  vanish by symmetry; this is 
evident since $\sum_{x=1}^{N}R_{i,x}=0$ for any $i=1,\ldots,N-1$. 
Similarly, the sequences like $a^{\dagger}g^{\dagger}g g$ also do not 
contribute. The sequences like $a^{\dagger}a^{\dagger}g g$ are 
diagonal in the orbital indices of the two $a^{\dagger}$ operators, since 
$\sum_{x=1}^{N}R_{i,x}R_{j,x}=\d_{ij}$; hence, they annihilate 
$|\Q_{0}(N,U)\ket$ (recall that in this state   all the $a$ 
orbitals are singly 
occupied, as we can see by direct inspection of the expressions for  
$|\F_{AF}^{0,0}\ket$ and $|\F_{AF}^{1,m}\ket$, $m=0,\pm 1$, of 
Section \ref{w=0states}). By a particle-hole transformation  
one can show that the remaining terms 
that create pairs in the shell ${\cal S}_{hf}$, like 
$a^{\dagger}a^{\dagger}g a$, etc., have no effect, as in 
Eqs.(\ref{mancava}-\ref{no2b}); we postpone 
the proof of this last spectacular cancellation until Appendix \ref{spr}. 
To sum up, no term in $W$ can change the number of particles in ${\cal 
S}_{hf}$, that is, that number is independent of $t$ and $U$! This proves the remarkable property that we have named 
{\it Shell Population Rigidity}: it holds 
for any finite $N$ and it is not specific of the thermodynamic limit. 
We shall see below that the {\it Shell Population Rigidity} 
characterizes the ground state and a suitable subspace of the total Hilbert
space.
In Fig.\ref{sprfig} we have drawn a physical picture of the 
cancellation among diagrams which do not preserve the number 
$n_{a}+n_{b}$. We can say that the $W=0$ states 
entering in our {\it Ansatz} are stable with respect to the Hubbard 
interaction $W$ also in the presence of other particles in the system.
This fact allows to exactly solve the Schr\"odinger 
equation. 
\begin{figure}[H]
\begin{center}
	\epsfig{figure=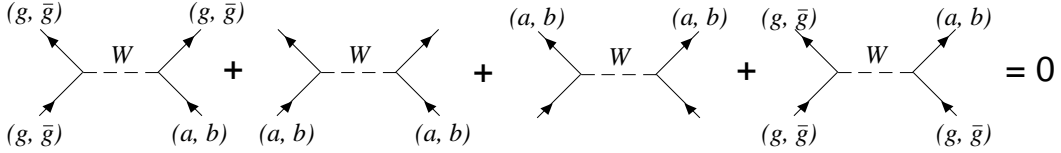,width=14cm}\caption{\footnotesize{
	Cancellation among the scattering amplitudes in the ansatz of 
	Equation (\ref{ansatz})  which do not preserve 
	the number of particles in the shell ${\cal S}_{hf}$. }}
\label{sprfig}
\end{center}
\end{figure}

Taking  these cancellations into account one obtains
\begin{eqnarray}
W_{{\cal A}}&:=&\frac{U}{4N}(g^{\dag}_{\ua}+\bar{g}^{\dag}_{\ua})
(g^{\dag}_{\da}+\bar{g}^{\dag}_{\da})(g_{\da}+\bar{g}_{\da})
(g_{\ua}+\bar{g}_{\ua})+\nonumber \\&+&
\frac{U}{2N}\sum_{i=1}^{N-1}
\left[a^{\dag}_{i\ua}(g^{\dag}_{\da}+\bar{g}^{\dag}_{\da})+
(g^{\dag}_{\ua}+\bar{g}^{\dag}_{\ua})a^{\dag}_{i\da}\right]
\left[(g_{\da}+\bar{g}_{\da})a_{i\ua}+
a_{i\da}(g_{\ua}+\bar{g}_{\ua})\right]
\end{eqnarray}
and
\begin{eqnarray}
W_{{\cal B}}&:=&\frac{U}{4N}(-g^{\dag}_{\ua}+\bar{g}^{\dag}_{\ua})
(-g^{\dag}_{\da}+\bar{g}^{\dag}_{\da})(-g_{\da}+\bar{g}_{\da})
(-g_{\ua}+\bar{g}_{\ua})+\nonumber \\&+&
\frac{U}{2N}\sum_{i=1}^{N-1}
\left[b^{\dag}_{i\ua}(-g^{\dag}_{\da}+\bar{g}^{\dag}_{\da})+
(-g^{\dag}_{\ua}+\bar{g}^{\dag}_{\ua})b^{\dag}_{i\da}\right]
\left[(-g_{\da}+\bar{g}_{\da})b_{i\ua}+
b_{i\da}(-g_{\ua}+\bar{g}_{\ua})\right]\;.
\end{eqnarray}

With $W$ written in the transformed picture one can calculate 
$W|\Q_{0}(N,U)\ket$. After very long, but 
standard algebra one finds
\begin{eqnarray}
W|g^{0}\ket\otimes|\F_{AF}^{0,0}\ket&=&U\left\{
\frac{2N-1}{2N}|g^{0}\ket\otimes|\F_{AF}^{0,0}\ket+
\frac{1}{2N}|\bar{g}^{0}\ket\otimes|\F_{AF}^{0,0}\ket+\right.
\nonumber \\ &+&
\frac{1}{2N}\sum_{i=1}^{N-1}|[g\bar{g}]^{1,1}\ket\otimes
(-S^{-a}_{i}+S^{-b}_{i})|\F_{AF}^{0,0}\ket+
\nonumber \\ &+&
\frac{1}{2N}\sum_{i=1}^{N-1}
\left[\bar{g}^{\dag}_{\ua}g^{\dag}_{\da}(\hat{n}^{a}_{i\da}-n^{b}_{i\da})+
\bar{g}^{\dag}_{\da}g^{\dag}_{\ua}(-\hat{n}^{a}_{i\ua}+n^{b}_{i\ua})
\right]|\F_{AF}^{0,0}\ket+
\nonumber \\ &+& \left.
\frac{1}{2N}\sum_{i=1}^{N-1}|[g\bar{g}]^{1,-1}\ket\otimes
(S^{+a}_{i}-S^{+b}_{i})|\F_{AF}^{0,0}\ket\right\}
\label{wg0}
\end{eqnarray}
and
\begin{eqnarray}
W|\bar{g}^{0}\ket\otimes|\F_{AF}^{0,0}\ket&=&U\left\{
\frac{2N-1}{2N}|\bar{g}^{0}\ket\otimes|\F_{AF}^{0,0}\ket+
\frac{1}{2N}|g^{0}\ket\otimes|\F_{AF}^{0,0}\ket+\right.
\nonumber \\ &+&
\frac{1}{2N}\sum_{i=1}^{N-1}|[g\bar{g}]^{1,1}\ket\otimes
(S^{-a}_{i}-S^{-b}_{i})|\F_{AF}^{0,0}\ket+
\nonumber \\ &+&
\frac{1}{2N}\sum_{i=1}^{N-1}
\left[g^{\dag}_{\da}\bar{g}^{\dag}_{\ua}(-\hat{n}^{a}_{i\ua}+n^{b}_{i\ua})+
g^{\dag}_{\ua}\bar{g}^{\dag}_{\da}(\hat{n}^{a}_{i\da}-n^{b}_{i\da})
\right]|\F_{AF}^{0,0}\ket+
\nonumber \\ &+& \left.
\frac{1}{2N}\sum_{i=1}^{N-1}|[g\bar{g}]^{1,-1}\ket\otimes
(-S^{+a}_{i}+S^{+b}_{i})|\F_{AF}^{0,0}\ket\right\}\;,
\label{wgbarra0}
\end{eqnarray}
while for the singlet 
$\sum_{m=-1}^{1}(-)^{m}|[g\bar{g}]^{1,m}\ket\otimes|\F_{AF}^{1,-m}\ket$ 
one obtains 
\begin{eqnarray}
W\sum_{m=-1}^{1}&(-)^{m}&|[g\bar{g}]^{1,m}\ket\otimes|\F_{AF}^{1,-m}\ket=
U\left\{\frac{N+1}{N}
\sum_{m=-1}^{1}(-)^{m}|[g\bar{g}]^{1,m}\ket\otimes|\F_{AF}^{1,-m}\ket+
\frac{1}{2N}(|g^{0}\ket-|\bar{g}^{0}\ket)\otimes\right.
\nonumber \\ &\otimes&\sum_{i=1}^{N-1}\left.
\left[
(S^{+a}_{i}-S^{+b}_{i})|\F_{AF}^{1,-1}\ket+
\frac{1}{\sqrt{2}}(n^{a}_{i\ua}-n^{a}_{i\da}-n^{b}_{i\ua}+n^{b}_{i\da})
|\F_{AF}^{1,0}\ket+(-S^{-a}_{i}+S^{-b}_{i})|\F_{AF}^{1,1}\ket\right]
\right\}\;.
\label{wggbarra1}
\end{eqnarray}

By means of the identities in Eqs.(\ref{id1}-\ref{id2}-\ref{id3}) one  
can write Eqs.(\ref{wg0}-\ref{wgbarra0}) in a compact form 
\begin{equation}
W|g^{0}\ket\otimes|\F_{AF}^{0,0}\ket=U\left\{
\frac{2N-1}{2N}|g^{0}\ket\otimes|\F_{AF}^{0,0}\ket+
\frac{1}{2N}|\bar{g}^{0}\ket\otimes|\F_{AF}^{0,0}\ket+
\frac{\sqrt{2}}{2N}\sum_{m=-1}^{1}
(-)^{m}|[g\bar{g}]^{1,m}\ket\otimes|\F_{AF}^{1,-m}\ket\right\}
\label{comp1}
\end{equation}
and
\begin{equation}
W|\bar{g}^{0}\ket\otimes|\F_{AF}^{0,0}\ket=U\left\{
\frac{2N-1}{2N}|\bar{g}^{0}\ket\otimes|\F_{AF}^{0,0}\ket+
\frac{1}{2N}|g^{0}\ket\otimes|\F_{AF}^{0,0}\ket-
\frac{\sqrt{2}}{2N}\sum_{m=-1}^{1}
(-)^{m}|[g\bar{g}]^{1,m}\ket\otimes|\F_{AF}^{1,-m}\ket\right\}\;.
\label{comp2}
\end{equation}

Let us now consider the second row in Eq.(\ref{wggbarra1}). If our 
{\it Ansatz} is correct, the state in the square brackets must be 
proportional to $|\F_{AF}^{0,0}\ket$. In this way one could close the 
equations and find an {\it exact} eigenstate. In Appendix \ref{proof} we 
prove that this is the case and in particular that 
\begin{equation}
\sum_{i=1}^{N-1}\left[
(S^{+a}_{i}-S^{+b}_{i})|\F_{AF}^{1,-1}\ket+
\frac{1}{\sqrt{2}}(n^{a}_{i\ua}-n^{a}_{i\da}-n^{b}_{i\ua}+n^{b}_{i\da})
|\F_{AF}^{1,0}\ket+(-S^{-a}_{i}+S^{-b}_{i})|\F_{AF}^{1,1}\ket\right]=
\sqrt{2}(N^{2}-1)|\F_{AF}^{0,0}\ket\;.
\label{ID4}
\end{equation}
Hence, Eq.(\ref{wggbarra1}) yields
\begin{eqnarray}
W\sum_{m=-1}^{1}(-)^{m}|[g\bar{g}]^{1,m}\ket\otimes|\F_{AF}^{1,-m}\ket=
U\left\{\frac{N+1}{N}
\sum_{m=-1}^{1}(-)^{m}|[g\bar{g}]^{1,m}\ket\otimes|\F_{AF}^{1,-m}\ket+
\right.
\nonumber \\ +\left.
\sqrt{2}\frac{N^{2}-1}{2N}(|g^{0}\ket-|\bar{g}^{0}\ket)\otimes
|\F_{AF}^{0,0}\ket\right\}\;.
\end{eqnarray}
This result, together with Eqs.(\ref{comp1}-\ref{comp2}) and 
Eqs.(\ref{kinact}), allows us to reduce the Schrodinger equation 
$(H_{{\mathrm Hubbard}}-E)|\Q_{0}(N,U)\ket=0$ to the diagonalization of the matrix
\begin{equation}
H(N,U)=\left(\begin{array}{ccc}
    -2t+\frac{2N-1}{2N}U      &          \frac{U}{2N}          & \frac{\sqrt{2(N^{2}-1)}}{2N}U  \\
    & & \\
       \frac{U}{2N}           &      2t+\frac{2N-1}{2N}U       & -\frac{\sqrt{2(N^{2}-1)}}{2N}U \\
       & & \\
\frac{\sqrt{2(N^{2}-1)}}{2N}U & -\frac{\sqrt{2(N^{2}-1)}}{2N}U & \frac{N+1}{N}U
\end{array}\right)\;,
\label{henne}
\end{equation}
where the relations
\begin{equation}\left[
\bra g^{0}|\otimes\bra\F_{AF}^{0,0}|\right]
\left[|g^{0}\ket\otimes|\F_{AF}^{0,0}\ket\right]=\left[
\bra \bar{g}^{0}|\otimes\bra\F_{AF}^{0,0}|\right]
\left[|\bar{g}^{0}\ket\otimes|\F_{AF}^{0,0}\ket\right]
\end{equation}
and
\begin{eqnarray}
\bra g^{0}|\otimes\bra\F_{AF}^{0,0}|W
\sum_{m=-1}^{1}(-)^{m}|[g\bar{g}]^{1,m}\ket\otimes|\F_{AF}^{1,-m}\ket=
\sum_{m=-1}^{1}(-)^{m}\bra\F_{AF}^{1,-m}|\otimes\bra[g\bar{g}]^{1,m}|
W|g^{0}\ket\otimes|\F_{AF}^{0,0}\ket,\nonumber \\
\bra \bar{g}^{0}|\otimes\bra\F_{AF}^{0,0}|W
\sum_{m=-1}^{1}(-)^{m}|[g\bar{g}]^{1,m}\ket\otimes|\F_{AF}^{1,-m}\ket=
\sum_{m=-1}^{1}(-)^{m}\bra\F_{AF}^{1,-m}|\otimes\bra[g\bar{g}]^{1,m}|
W|\bar{g}^{0}\ket\otimes|\F_{AF}^{0,0}\ket
\label{normrel}
\end{eqnarray} 
have been used to express 
$\sum_{m=-1}^{1}\bra\F_{AF}^{1,m}|\F_{AF}^{1,m}\ket$ in terms of 
$\bra\F_{AF}^{0,0}|\F_{AF}^{0,0}\ket$. 

The thermodynamic limit $N\ra\inf$ is well defined and non-trivial if 
we rescale $\tilde{t}$ in such a way that $N\tilde{t}=t=$const (which 
corresponds to the case of a fixed energy gap between the first two 
energy levels).  
The Hamiltonian matrix in Eq.(\ref{henne}) becomes 
\begin{equation}
H(\inf,U)=\left(\begin{array}{ccc}
-2t+U & 0 & U/\sqrt{2} \\
0 & 2t+U & -U/\sqrt{2} \\
U/\sqrt{2} & -U/\sqrt{2} & U
\end{array}\right)
\end{equation}
with eigenvalues
\begin{equation}
{\cal E}_{0}=U,\;\;\;{\cal E}_{\pm}=U\pm\D,
\;\;\;\;\D\equiv\sqrt{U^{2}+4t^{2}}\;.
\label{thermene}
\end{equation}
The (unnormalized) eigenvector corresponding to the lowest eigenvalue ${\cal 
E}_{-}$ gives 
\begin{equation}
|\Q_{0}(\inf,U)\ket=\left[\left(\frac{U}{t}-\frac{2\D}{U}(2+\frac{\D}{t})\right)|g^{0}\ket+
\frac{U}{t}|\bar{g}^{0}\ket\right]
\otimes|\F_{AF}^{0,0}\ket+
\sqrt{2}(2+\frac{\D}{t})\sum_{m=-1}^{1}(-)^{m}|[g\bar{g}]^{1,m}\ket\otimes|\F_{AF}^{1,-m}\ket
\label{intgs}\;.
\end{equation}

So far we have proved that the {\it Ansatz} in Eq.(\ref{ansatz}) 
gives three {\it exact} eigenstates of the CBG-Hubbard model 
at half filling. The eigenvalue of lowest 
energy ${\cal E}_{-}$ reduces to 
the non-interacting ground state energy for $U\ra 0$, while the 
associated eigenstate (\ref{intgs}) to the 
state $|g^{0}\ket\otimes|\F_{AF}^{0,0}\ket$. On the other hand, 
when $U\ra\inf$ the original Hamiltonian can be mapped onto what we 
can define the {\em   
CBG-Heisenberg Hamiltonian}:
\begin{equation}
H_{{\mathrm Heisenberg}}=\frac{4\tilde{t}^{2}}{U}
\sum_{x\in{\cal A}}\sum_{y\in{\cal B}}
({\mathbf S}_{x}\cdot{\mathbf S}_{y}-\frac{1}{4})\;.
\end{equation}
This model is exactly solvable and the ground state is obtained by
projecting onto the singlet the state where $N$ particles lie on 
the ${\cal A}$-sites with spin up and the remaining $N$ on the 
${\cal B}$-sites with spin down (Neel state): one finds
\begin{equation}
|\Q_{0}(N,\inf)\ket=\sum_{P}(-)^{P}\prod_{x=1}^{N}\s^{\dag}_{x,P(x+N)}|0\ket
\label{gspnheis}
\end{equation}
with $\s^{\dag}_{x,y}=\frac{1}{\sqrt{2}}(c^{\dag}_{x\ua}c^{\dag}_{y\da}
-c^{\dag}_{x\da}c^{\dag}_{y\ua})$. The ground state energy is 
$-2t^{2}/U$ for $N\ra \inf$ and is equal to the first order approximation of ${\cal 
E}_{-}$ in the small parameter $(t/U)^{2}$ (the same conclusion holds for 
any finite $N$, although it is more tedious to prove). 
Furthermore, by direct inspection of the lowest energy eigenvector of 
$H(N,U)$ one can show that $|\Q_{0}(N,U)\ket$ becomes the state 
in Eq.(\ref{gspnheis}) when $U\ra\inf$. Therefore, $|\Q_{0}(N,U)\ket$ reduces 
to the ground state of the CBG-Hubbard 
model in the two opposite limits $U\ra 0$ and $U\ra\inf$. 

To prove that the lowest energy eigenstate of $H(N,U)$ is the unique 
ground state we are looking for, one can exploit the ground state 
uniqueness of the Heisenberg Hamiltonian proved in 
Ref.\cite{liebmattis}. Since $|\Q_{0}(N,\inf)\ket$  is the 
ground state of the Heisenberg model and since the ground state of 
the half-filled Hubbard model is unique, a level crossing for some value 
of $U$ would be required if $|\Q_{0}(N,U)\ket$ were an 
excited state, contradicting the uniqueness. 

In conclusion we have proved that the {\it Ansatz} in 
Eq.(\ref{ansatz}) with $(\g_{g},\g_{\bar{g}},\g_{0})$ given by the 
lowest energy eigenvector of the matrix in Eq.(\ref{henne}) is the 
exact ground state of the half-filled Hubbard model defined on the 
CBG of size $|\L|=2N$ and repulsion $U$. In the 
next Section we shall calculate some physical quantities, as the 
spin-spin correlation function, and we shall prove that the particles are 
{\it antiferromagnetically} ordered. 

\section{Results and Discussion}
\label{results}

The exact ground state solution for arbitrary but finite $N$ allows 
to study the ground state energy $E_{0}(N,U)$ as a function of $N$ and $U$. 
Taking $\tilde{t}N=t=1$, in Fig.(\ref{energy}) we have plotted 
$E_{0}(N,U)$ in the range $N=1,..,10$ and $0<U<20$. 
\begin{figure}[H]
\begin{center}
	\epsfig{figure=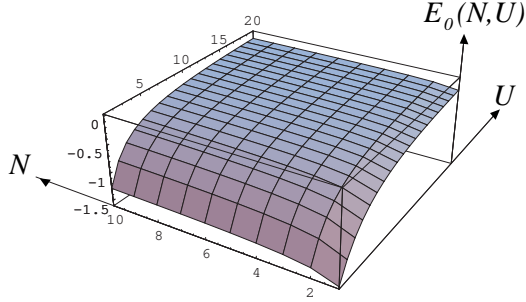,width=9cm}\caption{\footnotesize{
	Ground state energy $E_{0}(N,U)$ of the half-filled Hubbard model on the 
	CBG as a function of the on-site repulsion 
	parameter $U$ and the number of lattice sites $N$ in each sublattice. 
	The hopping parameter has been chosen to be $t=1$.}}
\label{energy}
\end{center}
\end{figure}
The figure shows 
that $E_{0}(N,U)$ is a monotonically increasing function of $N$ and 
$U$, that is $E_{0}(N+1,U)-E_{0}(N)>0$ and $\de E_{0}(N,U)/\de U>0$. 
The limit $N\ra \inf$ at fixed $U$ is given by ${\cal E}_{-}$ in 
Eq.(\ref{thermene}):
\begin{equation}
\lim_{|\L|\ra\inf}\lim_{\b\ra\inf}
E^{(\b)}(|\L|,t,U)=U-\sqrt{U^{2}+4t^{2}}<0
\label{groundex}
\end{equation}
where $E^{(\b)}(|\L|,t,U)$ is the internal energy of the system at 
inverse Temperature $\b$ and number of lattice sites $|\L|=2N$. On the 
other hand, one can calculate the internal energy at zero Temperature in 
thermodynamic limit: $E^{(\b)}(|\L|\ra\inf,t,U)=\W_{\b}+k_{B}S/\b+
\m |\L| n$ with $\W_{\b}$ from Eq.(\ref{therm2}) and $S$ the entropy.  
Following the procedure of van Dongen and Vollhardt one gets
\begin{equation}
\lim_{\b\ra\inf}\lim_{|\L|\ra\inf}
E^{(\b)}(|\L|,t,U)=\left\{\begin{array}{ll}
-2t=-2N\tilde{t} <0 & {\mathrm if}\;\tilde{t}\;{\mathrm is} \;
{\mathrm not}\; {\mathrm rescaled}\\
O(1) & {\mathrm if}\;\tilde{t}\sim 1/|\L|
\end{array}\right.\;.
\label{groundth}
\end{equation}
In the case $\tilde{t}$=const  
the two limits commute and the trivial thermodynamic behavior 
of the model originates from the infinite gap between the 
lowest level and the zero-energy levels: the two particles in the $g$ 
orbitals are completely decoupled from the dynamics of the system. 
On the other hand, if $\tilde{t}$ is rescaled the gap remains 
frozen and taking $|\L|\ra\inf$ first, we can only say that the ground state 
energy is $O(1)$, but we cannot predict the exact amount within the 
scheme proposed by van Dongen and Vollhardt. This is a consequence of the fact 
that in the limit $|\L|\ra\inf$, we retain only the extensive contributions to the 
internal energy. 

The ground state energy 
Eq.(\ref{groundex}) is always higher than that of the non-interacting 
case and hence non-trivial correlations survive in the thermodynamic 
limit. We have calculated the 
ground state average of the number of doubly occupied sites: 
\begin{eqnarray}
\overline{D}(N,U)\equiv\frac{\overline{W}(N,U)}{U}\equiv\frac{1}{U}
\bra\Q_{0}(N,U)|W|\Q_{0}(N,U)\ket&=&\frac{1}{U}
\bra\Q_{0}(N,U)|H_{{\mathrm Hubbard}}-H_{0}|\Q_{0}(N,U)\ket=
\nonumber \\ &=& \frac{1}{U}
E_{0}(N,U)+\frac{2t}{U}[\g^{2}_{g}(N,U)-\g^{2}_{{\bar g}}(N,U)]
\end{eqnarray}
where $\g_{g}$ and $\g_{{\bar g}}$ are the first two components of 
the normalized ground state vector, see Eq.(\ref{ansatz}). In 
Fig.(\ref{deltaw}$a$) we have plotted the trend of $\overline{D}(N,U)$ as 
$N$ increases for different values of $U$. As expected 
$\overline{D}(N,U)$ is a monotonically decreasing function of $U$,  
$\de \overline{D}(N,U)/\de U<0$, approaching zero for $U\ra\inf$ [where
the exact ground state reduces to the antiferromagnetic Neel state, 
see Eq.(\ref{gspnheis})]. Nevertheless, $\overline{D}(N,U)$ shows a 
non-trivial behaviour as $N$ increases at fixed $U$ values. In 
the weak coupling regime, $U\ll t$, the number of doubly occupied sites 
grows as $N$ becomes larger and larger converging to a finite 
value when $N\ra\inf$. The opposite trend is observed in the strong 
coupling regime, $U\gg t$, where $\overline{D}(N,U)$ decreases as $N$ 
increases. In the intermediate regime $U\sim t$, $\overline{D}(N,U)$ 
is an increasing function of $N$ for small $N$, but becomes a 
decreasing function for large $N$. 
Hence, there is a critical value $U_{c}(N)$ where the analytic 
continuation of  $\overline{D}(N,U)$ to real $N$ verifies 
\begin{equation}
\left(\frac{\de^{2}\overline{D}}{\de U\de N}\right)_{U_{c}(N)}=0
\end{equation}
\begin{figure}[H]
\begin{center}
	\epsfig{figure=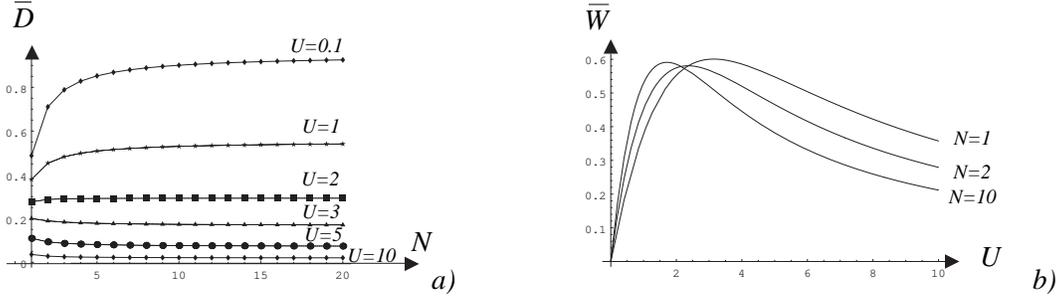,width=14cm}\caption{\footnotesize{
        $a$) Trend of the ground state average of the number of 
        doubly occupied sites versus $N$ in the range $1\leq N\leq 20$ 
        for different values of the Hubbard interaction parameter 
        $U=$0.1, 1, 2, 3, 5, 10. $b$) Ground state average of the 
        interaction term $W$ as a function of $U$ in the range $0\leq 
        U\leq 10$ for three different number of sublattice sites, 
        $N=$ 1, 2, 10.
	The hopping parameter has been chosen to be $t=1$ in both cases.}}
\label{deltaw}
\end{center}
\end{figure}
In Fig.(\ref{deltaw}$b$) we report the ground state average of the 
interacting term $W$ as a function of $U$ for three different values 
of $N$. There is always a critical value for $U$ where the repulsion 
is maximized. Moreover, $\lim_{N\ra\inf} \overline{W}(N,U)\equiv 
\overline{W}_{\inf}(U)\neq 0$ for any finite $U$ and the system 
cannot avoid double occupation neither in the thermodynamic limit. 
In the limit $U\ra\inf$, $\overline{W}(N,U)\sim \frac{1}{U}$, that 
is,  $\overline{W}(N,U)\ra 0$ in agreement with the 
Gutzwiller {\it Ansatz}\cite{gutz}.

Next, we have calculated the spin-spin correlation function
\begin{equation}
G_{{\mathrm spin}}(x,y)\equiv 
\bra\Q_{0}(N,U)|S^{z}_{x}S^{z}_{y}|\Q_{0}(N,U)\ket=
\lim_{\b\ra\inf}G_{{\mathrm spin}}(x,y;\b,h=0,|\L|)=
\lim_{\b\ra\inf}\lim_{h\ra 0}G_{{\mathrm spin}}(x,y;\b,h,|\L|)
\end{equation}
where $G_{{\mathrm spin}}(x,y;\b,h,|\L|)$ was defined in Section 
\ref{thermolimi}, $S^{z}_{x}=\frac{1}{2}(\hat{n}_{x\ua}-\hat{n}_{x\da})$ 
and $|\Q_{0}\ket$ is the normalized ground state 
\begin{equation}
|\Q_{0}\ket=\left[\frac{\g_{g}}{{\cal N}_{0}}|g^{0}\ket+
\frac{\g_{\bar{g}}}{{\cal N}_{0}}|\bar{g}^{0}\ket\right]
\otimes|\F_{AF}^{0,0}\ket+\frac{1}{\sqrt{3}}
\frac{\g_{0}}{{\cal N}_{1}}\sum_{m=-1}^{1}(-)^{m}|[g\bar{g}]^{1,m}
\ket\otimes|\F_{AF}^{1,-m}\ket\;,
\label{ansatznorm}
\end{equation}
with ${\cal N}^{2}_{0}=\bra\F_{AF}^{0,0}|\F_{AF}^{0,0}\ket$, 
${\cal N}^{2}_{1}=\bra\F_{AF}^{1,m}|\F_{AF}^{1,m}\ket$, $m=0,\pm 1$,  
and $\g_{g}^{2}+\g_{\bar{g}}^{2}+\g_{0}^{2}=1$. 
Due to the $S_{N}\otimes S_{N}\otimes Z_{2}$ symmetry, $G_{{\mathrm 
spin}}(x,y)$ can be written as 
\begin{equation}
G_{{\mathrm spin}}(x,y)=\left\{\begin{array}{ll}
G_{0} & x=y \\
G_{{\mathrm on}} & x\in{\cal A}\;(x\in{\cal B})\;{\mathrm and}\;
y\in{\cal A}\;(y\in{\cal B})\\
G_{{\mathrm off}} & x\in{\cal A}\;(x\in{\cal B})\;{\mathrm and}\;
y\in{\cal B}\;(x\in{\cal A})
\end{array}\right.
\end{equation}
and by exploiting the sum rule $\sum_{y\in\L}G_{{\mathrm spin}}(x,y)=
G_{0}+(N-1)G_{{\mathrm on}} +NG_{{\mathrm off}}=0$, 
one can express $G_{{\mathrm on}}$ in 
terms of $G_{{\mathrm off}}$ and $G_{0}$: 
\begin{equation}
G_{{\mathrm on}}=-\frac{N}{N-1}G_{{\mathrm off}}-
\frac{G_{0}}{N-1}\;.
\end{equation}
The problem is then reduced to the evaluation of $G_{0}$ and $G_{{\mathrm 
off}}$. We have 
\begin{eqnarray}
G_{0}&=&G_{{\mathrm spin}}(x,x)=\bra\Q_{0}|(S^{z}_{x})^{2}|\Q_{0}\ket=
\frac{1}{2N}\sum_{x\in\L}\bra\Q_{0}|(S^{z}_{x})^{2}|\Q_{0}\ket=
\nonumber \\ &=& 
\frac{1}{8N}\sum_{x\in\L}\bra\Q_{0}|\hat{n}_{x\ua}+\hat{n}_{x\da}-2
\hat{n}_{x\ua}\hat{n}_{x\da}|\Q_{0}\ket=
\frac{1}{4}(1-\frac{\overline{D}}{N})\;.
\end{eqnarray}
More effort is needed to calculate $G_{{\mathrm off}}$. Expanding 
the $c$ operators in $g$, ${\bar g}$, $a$ and $b$ operators of 
Eq.(\ref{traop}) and taking into account that 
$(\sum_{x\in\L}S^{z}_{x})^{2}|\Q_{0}\ket=0$, we get
\begin{eqnarray}
G_{{\mathrm off}}=-\frac{1}{N^{2}}\sum_{x\in{\cal A}}\sum_{y\in{\cal A}}
\bra\Q_{0}|S^{z}_{x}S^{z}_{y}|\Q_{0}\ket=-\frac{1}{N^{2}}
\left[\frac{1}{16}\bra\Q_{0}|\left(\sum_{\s}(-)^{\s}(g^{\dag}_{\s}+\bar{g}^{\dag}_{\s})
(g_{\s}+\bar{g}_{\s})\right)^{2}|\Q_{0}\ket\right.+
\nonumber \\ +\left.\frac{1}{2}
\bra\Q_{0}|S^{za}\sum_{\s}(-)^{\s}(g^{\dag}_{\s}+\bar{g}^{\dag}_{\s})
(g_{\s}+\bar{g}_{\s})|\Q_{0}\ket+
\bra\Q_{0}|S^{za}S^{za}|\Q_{0}\ket\right]
\end{eqnarray}
with $(-)^{\s}=+,-$ for $\s=\ua,\da$. After long but standard algebra one obtains 
\begin{eqnarray}
G_{{\mathrm off}}=-\frac{1}{N^{2}}\left\{
\frac{1}{8}\left[(\g_{g}-\g_{{\bar g}})^{2}+2\g_{0}^{2}\right]-
\sqrt{\frac{2}{3}}\frac{\g_{0}(\g_{g}-\g_{{\bar g}})}{{\cal N}_{0}
{\cal N}_{1}}\bra \F_{AF}^{1,0}|S^{za}|\F_{AF}^{0,0}\ket+
\frac{\g_{g}^{2}+\g^{2}_{{\bar g}}}{{\cal N}_{0}^{2}}
\bra \F_{AF}^{0,0}|S^{za}S^{za}|\F_{AF}^{0,0}\ket+
\right.
\nonumber \\ +\left.
\frac{\g_{0}^{2}}{3{\cal N}_{1}^{2}}\left[
\bra \F_{AF}^{1,-1}|S^{za}|\F_{AF}^{1,-1}\ket-
\bra \F_{AF}^{1,1}|S^{za}|\F_{AF}^{1,1}\ket+
\sum_{m=-1}^{1}\bra \F_{AF}^{1,m}|S^{za}S^{za}|\F_{AF}^{1,m}\ket
\right]\right\}\;.
\label{quasigoff}
\end{eqnarray}
and exploiting the $Z_{2}$ symmetry $a_{i}\ra b_{i}$ and $b_{i}\ra 
a_{i}$, which implies $|\F_{AF}^{S,M_{S}}\ket\ra (-)^{S}
|\F_{AF}^{S,M_{S}}\ket$, 
\begin{equation}
\bra \F_{AF}^{1,0}|S^{za}|\F_{AF}^{0,0}\ket=
-\bra \F_{AF}^{1,0}|S^{zb}|\F_{AF}^{0,0}\ket=
\frac{1}{2}\bra \F_{AF}^{1,0}|S^{za}-S^{zb}|\F_{AF}^{0,0}\ket=
\frac{1}{2}{\cal N}_{1}^{2}
\label{rich}
\end{equation}
\begin{equation}
\bra \F_{AF}^{1,1}|S^{za}|\F_{AF}^{1,1}\ket=\frac{1}{2}
\bra \F_{AF}^{1,1}|S^{za}+S^{zb}|\F_{AF}^{1,1}\ket=
\frac{1}{2}{\cal N}_{1}^{2}
\label{rich2}
\end{equation}
\begin{equation}
\bra \F_{AF}^{1,-1}|S^{za}|\F_{AF}^{1,-1}\ket=\frac{1}{2}
\bra \F_{AF}^{1,-1}|S^{za}+S^{zb}|\F_{AF}^{1,-1}\ket=
-\frac{1}{2}{\cal N}_{1}^{2}
\label{rich3}
\end{equation}
where in Eq.(\ref{rich}) we have used the identity Eq.(\ref{id2}). 
Finally we recall that the $W=0$ states $|\F_{AF}^{S,M_{S}}\ket$ are 
eigenstates of the square of the 
total spin operators of each sublattice 
$S^{2}_{{\cal A}}\equiv 
(S^{za})^{2}+\frac{1}{2}(S^{+a}S^{-a}+S^{-a}S^{+a})$ and 
$S^{2}_{{\cal B}}\equiv 
(S^{zb})^{2}+\frac{1}{2}(S^{+b}S^{-b}+S^{-b}S^{+b})$ with eigenvalue 
$(\frac{N-1}{2})(\frac{N-1}{2}+1)=\frac{1}{4}(N^{2}-1)$; hence
\begin{equation}
\sum_{M_{S}=-S}^{S}\bra\F_{AF}^{S,M_{S}}|(S^{za})^{2}
|\F_{AF}^{S,M_{S}}\ket=\frac{N^{2}-1}{12}(2S+1){\cal N}^{2}_{S};
\;\;\;\;{\cal 
N}^{2}_{S}=\bra\F_{AF}^{S,M_{S}}|\F_{AF}^{S,M_{S}}\ket,\;\;
M_{S}=-S,\ldots,S\;.
\label{angid}
\end{equation}
Substitution of Eqs.(\ref{rich}-\ref{rich2}-\ref{rich3}-\ref{angid}) in 
Eq.(\ref{quasigoff}) yields
\begin{equation}
G_{{\mathrm off}}=-\frac{1}{N^{2}}\left\{
\frac{N^{2}-1}{12}+\frac{1}{8}[(\g_{g}-\g_{{\bar g}})^{2}+2\g_{0}^{2}]
-\frac{1}{3}\g_{0}^{2}-\sqrt{\frac{N^{2}-1}{18}}\g_{0}(\g_{g}-\g_{{\bar g}})
\right\}
\end{equation}
where we have taken into account Eq.(\ref{normrel}) to evaluate the ratio 
${\cal N}_{1}/{\cal N}_{0}=\sqrt{(N^{2}-1)/3}$.
\begin{figure}[H]
\begin{center}
	\epsfig{figure=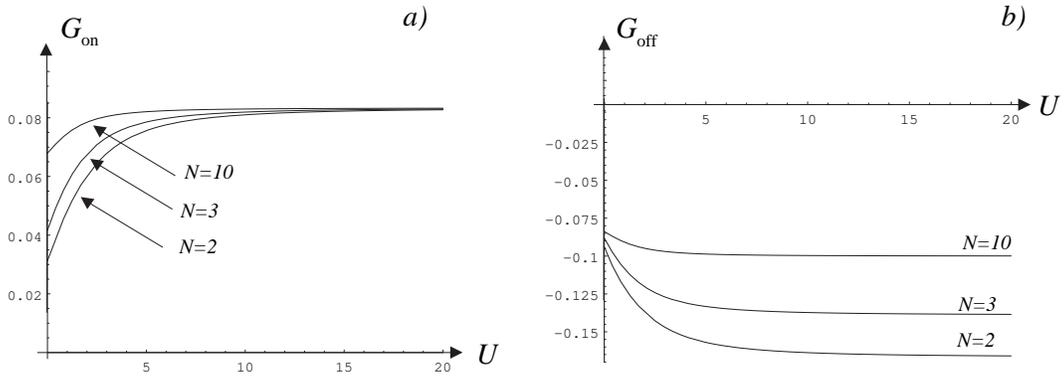,width=14cm}\caption{\footnotesize{
        $a$) $G_{{\mathrm on}}$ versus $U$ in the range $0\leq U\leq 
        20$ for three different values of the number of sites 
        $N=2,\;3,\;10$. $b$) $G_{{\mathrm off}}$ versus $U$ in the range 
	$0\leq U\leq 20$ for three different values of the number of sites 
        $N=2,\;3,\;10$.
	The hopping parameter has been chosen to be $t=1$ in both cases.}}
\label{gongoff}
\end{center}
\end{figure}

In Fig.(\ref{gongoff}) we report the trend of $G_{{\mathrm on}}$ and 
$G_{{\mathrm off}}$ versus $U$ for three different values of 
$N=2,\;3,\;10$. According to the Shen-Qiu-Tian theorem\cite{sqt}, 
$G_{{\mathrm on}}$ is always larger than zero while $G_{{\mathrm 
off}}$ is always negative. Now consider the ground state average 
of the square of the staggered magnetization operator
\begin{equation}
m^{2}_{AF}\equiv
\frac{1}{|\L|}\bra\Q_{0}|[\sum_{x\in\L}\e(x)S^{z}_{x}]^{2}
|\Q_{0}\ket,\;\;\;\;
\e(x)=1,\;-1\;{\mathrm for}\;\; x\in{\cal A},\;{\cal B}\;.
\label{sqtfor}
\end{equation}
The Shen-Qiu-Tian theorem implies that each term in the expansion 
of Eq.(\ref{sqtfor}) is non-negative. We emphasize, however, that 
for $|{\cal A}|=|{\cal B}|$ this does not imply that $m^{2}_{AF}$ 
is an extensive quantity! Remarkably, 
our results show that $m^{2}_{AF}=G_{0}+(N-1)G_{{\mathrm on}}-
NG_{{\mathrm off}}$ is extensive for any value of the on-site 
repulsion parameter $U$ and constitute the {\it first} example of antiferromagnetism in 
the ground state of an itinerant-electrons model of arbitrary size. 

We also observe that the thermodynamic limit yields a non-vanishing 
result:
\begin{equation}
\lim_{|\L|\ra\inf}G_{0}=\frac{1}{4},\;\;\;\;
\lim_{|\L|\ra\inf}G_{{\mathrm on}}=-
\lim_{|\L|\ra\inf}G_{{\mathrm off}}=\frac{1}{12},\;\;\;\;
\end{equation}
or
\begin{equation}
\lim_{|\L|\ra\inf}\lim_{\b\ra \inf}\lim_{h\ra 0}
|G_{{\mathrm spin}}(x,y;\b,h,|\L|)|=\frac{1}{12}\;\;\; {\mathrm for}\; 
x\neq y\;.
\label{correx}
\end{equation}
By comparing Eq.(\ref{correx}) with Eq.(\ref{corr}), we see that 
the $\lim_{\b\ra \inf}\lim_{h\ra 0}$ do not commute with the 
$\lim_{|\L|\ra\inf}$.

\section{Summary and Conclusions}
\label{summary}

We have introduced the CBG-Hubbard model and  explicitly written the 
ground state wave function at half filling for any repulsion parameter $U$. 
The topology of the graph $\L$ allows $(2N-2)$-body eigenstates 
$|\F_{AF}^{S,M_{S}}\ket$ of the Hamiltonian which are free of double occupation. 
They can be obtained projecting  the determinantal state $|\F_{AF}\ket$ on a 
given spin-$S$ subspace. $|\F_{AF}\ket$ has the {\it antiferromagnetic property}, 
that is the map ${\cal A}\leftrightarrow {\cal B}$ is equivalent to a spin-flip. 

We have shown that the ground state has a fixed  number of particles in the 
zero-energy one-body shell ${\cal S}_{hf}$, independent of $t$ and $U$. This very 
remarkable  {\it Shell Population Rigidity} holds in any finite-size 
system and it provides the closure of the equations. This implies 
extra-conservation laws in a suitable subspace of the full Hilbert 
space including the ground state. Only the singlet 
$|\F_{AF}^{0,0}\ket$ and the triplet 
$|\F_{AF}^{1,m}\ket$, $m=0,\pm 1$, are involved in the ground state expansion. 
Qualitatively, we may say that according to Eq.(\ref{intgs}) the particles 
in the shell ${\cal S}_{hf}$ manage to  avoid the double occupation.

We have calculated the spin-spin correlation function and shown that 
the ground state exhibits an antiferromagnetic order for any 
non-zero $U$ even in the thermodynamic limit. The kinetic term 
induces non-trivial correlations among the particles and an antiparallel 
spin configuration in the two sublattices turns out to be energetically  
favoured. Therefore, the model contains the basic ingredients to 
understand how the subtle competition between the delocalization induced 
by $H_{0}$ and the localization induced by $W$ may give rise to a 
magnetically ordered state even outside the strong coupling regime 
(where the Hubbard model is equivalent to the Heisenberg model). 
On the other hand, within the scheme proposed by van Dongen 
and Vollhardt, the model is well described by 
a paramagnetic Hamiltonian; the difference stems from taking the 
thermodynamic limit before the limit of zero 
Temperature, because then the kinetic term completly decouples 
from the interaction and loses any r$\hat{o}$le in the physical 
response functions. 

The present formalism lends itself to solve more realistic Hubbard Hamiltonians 
with an increasing number of negative and positive energy levels.
Higher-spin projections 
$|\F_{AF}^{S,M_{S}}\ket$ are involved in the ground state expansions; 
the results will be published elsewhere. We are currently 
investigating group-theory aspects of this model and its extensions.

Finally, we recall\cite{ssc2001}\cite{jop2001}\cite{jop2002} that 
the standard  Hubbard model on a $N\times N$ 
square lattice and periodic boundary conditions  also  has  
$|\F_{AF}^{S,M_{S}}\ket$-like eigenstates. It could be that the 
{\it Shell Population Rigidity} 
holds in this case too, but the proof is lacking. 

\appendix

\section{Shell Population Rigidity}
\label{spr}

We have seen in Section \ref{ground} that the most part of the 
contributions which do not preserve the number of particles in the 
shell ${\cal S}_{hf}$ yield nothing. To prove the {\it Shell 
Popoulation Rigidity} property we need to show that the sequences 
like $a^{\dag}a^{\dag}ga$ and similar terms with $b$ instead of $a$ 
and/or $g$ replaced by  $\bar{g}$ vanish or in formul\ae
\begin{equation}
\sum_{x\in{\cal A}}\sum_{ijm}R_{i,x}R_{j,x}R_{m,x}a^{\dag}_{i\ua}
a^{\dag}_{j\da}a_{m\s}|\F_{AF}^{S,M_{S}}\ket=0
\label{spr1}
\end{equation}
and the like with $a$ replaced by $b$. This last spectacular 
cancellation can be seen by particle-hole transforming 
Eq.(\ref{spr1}). Under a particle-hole 
transformation $c_{x\s}\ra c^{\dag}_{x\s}$ and the $W=0$ states 
$|\F_{AF}^{S,M_{S}}\ket\ra 
|g^{0}\bar{g}^{0}\ket\otimes|\F_{AF}^{S,-M_{S}}\ket$, 
modulo a phase factor. Hence, Eq.(\ref{spr1}) is equivalent to 
\begin{equation}
\sum_{x\in{\cal A}}\sum_{ijm}R_{i,x}R_{j,x}R_{m,x}a_{i\ua}
a_{j\da}a^{\dag}_{m\s}|\F_{AF}^{S,M_{S}}\ket=0\;.
\label{spr2}
\end{equation}
Let the spin of the operator $a^{\dag}_{m\s}$ be up for the sake of 
definiteness (the same reasoning holds in the case of down spin). The 
l.h.s. of Eq.(\ref{spr2}) can be rewritten as 
\begin{equation}
-\sum_{x\in{\cal A}}\sum_{ijm}R_{i,x}R_{j,x}R_{m,x}
[\d_{i,m}-a^{\dag}_{m\ua}a_{i\ua}]a_{j\da}|\F_{AF}^{S,M_{S}}\ket=
-\sum_{j}\left[\sum_{x}\sum_{i}R^{2}_{i,x}R_{j,x}\right]
a_{j\da}|\F_{AF}^{S,M_{S}}\ket
\end{equation}
where we have exploited the lack of double occupation of the $W=0$ state, see 
Eq.(\ref{no2a}). Therefore, we are left to prove that 
\begin{equation}
\sum_{x=1}^{N}\sum_{i=1}^{N-1}R_{j,x}R^{2}_{i,x}=0,\;\;\;\forall j\;.
\label{spr3}
\end{equation}
We recall that $R$ is an $(N-1)\times N$ rectangular matrix whose rows 
are orthonormal vectors which are orthogonal to the $N-$dimensional vector 
$(1,1,\ldots,1,1)$. It is a simple exercise to verify that 
\begin{equation}
R=\left(\begin{array}{cccccc}
\frac{1}{\sqrt{2}} & -\frac{1}{\sqrt{2}} & 0 & 0 & \ldots & 0 \\
\frac{1}{\sqrt{6}} &  \frac{1}{\sqrt{6}} & -\frac{2}{\sqrt{3}} & 0 & 
\ldots & 0 \\
: & : & : & : & : & : \\
\frac{1}{\sqrt{N(N+1)}} & \frac{1}{\sqrt{N(N+1)}} & 
\frac{1}{\sqrt{N(N+1)}} & \frac{1}{\sqrt{N(N+1)}} &  \ldots & -
\sqrt{\frac{N}{(N+1)}} 
\end{array}\right)\;\Rightarrow 
R_{m,x}=\left\{\begin{array}{ll}
0 & x> m+1 \\
-\sqrt{\frac{m}{m+1}} & x=m+1 \\
\frac{1}{m}\sqrt{\frac{m}{m+1}} & x\leq m 
\end{array}\right.
\end{equation}
is a correct choice and that Eq.(\ref{spr3}) is identically verified.

\section{Proof of Eq.(\ref{ID4})}
\label{proof}

To prove Eq.(\ref{ID4}) we shall use the definitions 
(\ref{trip0af}-\ref{trip1af}-\ref{trip-1af}) for the triplet $W=0$ state 
$|\F_{AF}^{1,m}\ket$, $m=0,\pm 1$. The first term yields 
\begin{eqnarray}
\sum_{k}(S^{+a}_{k}-S^{+b}_{k})|\F_{AF}^{1,-1}\ket&=&
\sum_{P}(-)^{P}\sum_{k}(S^{+a}_{k}-S^{+b}_{P(k)})
\sum_{j}\t^{(-1)^{\dag}}_{j,P(j)}\prod_{i\neq 
j}\s^{\dag}_{i,P(i)}|0\ket=
\nonumber \\ &=&
\sum_{P}(-)^{P}\sum_{j}(S^{+a}_{j}-S^{+b}_{P(j)})
\t^{(-1)^{\dag}}_{j,P(j)}\prod_{i\neq 
j}\s^{\dag}_{i,P(i)}|0\ket+
\nonumber \\ &+&
\sum_{P}(-)^{P}\sum_{j}\t^{(-1)^{\dag}}_{j,P(j)}
\sum_{k\neq j}(S^{+a}_{k}-S^{+b}_{P(k)})\s^{\dag}_{k,P(k)}
\prod_{i\neq 
j,k}\s^{\dag}_{i,P(i)}|0\ket=
\nonumber \\ &=&
\sqrt{2}\sum_{P}(-)^{P}\sum_{j}\s^{\dag}_{j,P(j)}
\prod_{i\neq 
j}\s^{\dag}_{i,P(i)}|0\ket+
\nonumber \\ &-&
\sqrt{2}\sum_{P}(-)^{P}\sum_{j}\sum_{k\neq j}\t^{(-1)^{\dag}}_{j,P(j)}
\t^{(+1)^{\dag}}_{k,P(k)}\prod_{i\neq 
j,k}\s^{\dag}_{i,P(i)}|0\ket=
\nonumber \\ &=&
\sqrt{2}(N-1)|\F_{AF}^{0,0}\ket-
\sqrt{2}\sum_{P}(-)^{P}\sum_{j}\sum_{k\neq j}\t^{(-1)^{\dag}}_{j,P(j)}
\t^{(+1)^{\dag}}_{k,P(k)}\prod_{i\neq 
j,k}\s^{\dag}_{i,P(i)}|0\ket\;,
\label{firstterm}
\end{eqnarray}
while second term yields
\begin{eqnarray}
\sum_{k}(\hat{n}^{a}_{k\ua}-\hat{n}^{a}_{k\da}&-&
\hat{n}^{b}_{k\ua}+\hat{n}^{b}_{k\da})|\F_{AF}^{1,0}\ket=
\sum_{P}(-)^{P}\sum_{k}(\hat{n}^{a}_{k\ua}-\hat{n}^{a}_{k\da}-
\hat{n}^{b}_{P(k)\ua}+\hat{n}^{b}_{P(k)\da})
\sum_{j}\t^{(0)^{\dag}}_{j,P(j)}\prod_{i\neq 
j}\s^{\dag}_{i,P(i)}|0\ket=
\nonumber \\ &=&
\sum_{P}(-)^{P}\sum_{j}(\hat{n}^{a}_{j\ua}-\hat{n}^{a}_{j\da}-
\hat{n}^{b}_{P(j)\ua}+\hat{n}^{b}_{P(j)\da})
\t^{(0)^{\dag}}_{j,P(j)}\prod_{i\neq 
j}\s^{\dag}_{i,P(i)}|0\ket+
\nonumber \\ &+&
\sum_{P}(-)^{P}\sum_{j}\t^{(0)^{\dag}}_{j,P(j)}
\sum_{k\neq j}(\hat{n}^{a}_{k\ua}-\hat{n}^{a}_{k\da}-
\hat{n}^{b}_{P(k)\ua}+\hat{n}^{b}_{P(k)\da})\s^{\dag}_{k,P(k)}
\prod_{i\neq 
j,k}\s^{\dag}_{i,P(i)}|0\ket=
\nonumber \\ &=&
2\sum_{P}(-)^{P}\sum_{j}\s^{\dag}_{j,P(j)}\prod_{i\neq j}
\s^{\dag}_{i,P(i)}|0\ket+
\nonumber \\ &+& 2
\sum_{P}(-)^{P}\sum_{j}\sum_{k\neq j}\t^{(0)^{\dag}}_{j,P(j)}
\t^{(0)^{\dag}}_{k,P(k)}
\prod_{i\neq 
j,k}\s^{\dag}_{i,P(i)}|0\ket=
\nonumber \\ &=&
2(N-1)|\F_{AF}^{0,0}\ket+
2\sum_{P}(-)^{P}\sum_{j}\sum_{k\neq j}\t^{(0)^{\dag}}_{j,P(j)}
\t^{(0)^{\dag}}_{k,P(k)}
\prod_{i\neq 
j,k}\s^{\dag}_{i,P(i)}|0\ket\;.
\label{secondterm}
\end{eqnarray}
The third term can be computed following the same steps of 
Eq.(\ref{firstterm}) and the final result is
\begin{equation}
\sum_{k}(-S^{-a}_{k}+S^{-b}_{k})|\F_{AF}^{1,1}\ket=
\sqrt{2}(N-1)|\F_{AF}^{0,0}\ket-\sqrt{2}
\sum_{P}(-)^{P}\sum_{j}\sum_{k\neq j}\t^{(+1)^{\dag}}_{j,P(j)}
\t^{(-1)^{\dag}}_{k,P(k)}\prod_{i\neq 
j,k}\s^{\dag}_{i,P(i)}|0\ket\;.
\label{thirdterm}
\end{equation}
Denoting by $[..]$ the left hand side of Eq.(\ref{ID4}), from 
Eqs.(\ref{firstterm}-\ref{secondterm}-\ref{thirdterm}) one obtains
\begin{eqnarray}
&&[..]=3\sqrt{2}(N-1)|\F_{AF}^{0,0}\ket-\sqrt{2}
\sum_{P}(-)^{P}\sum_{j}\sum_{k\neq j}
\left[\t^{(-1)^{\dag}}_{j,P(j)}\t^{(+1)^{\dag}}_{k,P(k)}-
\t^{(0)^{\dag}}_{j,P(j)}\t^{(0)^{\dag}}_{k,P(k)}+
\t^{(+1)^{\dag}}_{j,P(j)}\t^{(-1)^{\dag}}_{k,P(k)}\right]\prod_{i\neq 
j,k}\s^{\dag}_{i,P(i)}|0\ket.
\nonumber \\&&
\label{sq}
\end{eqnarray}
Let us consider the first and the third term in the square bracket of 
the right hand side of Eq.(\ref{sq}). We have
\begin{eqnarray}
\sum_{P'}(-)^{P'}
\left[\t^{(-1)^{\dag}}_{j,P'(j)}\t^{(+1)^{\dag}}_{k,P'(k)}+
\t^{(+1)^{\dag}}_{j,P'(j)}\t^{(-1)^{\dag}}_{k,P'(k)}\right]\prod_{i\neq 
j,k}\s^{\dag}_{i,P'(i)}|0\ket=\nonumber \\ =
-\sum_{P}(-)^{P}\left[\t^{(-1)^{\dag}}_{j,P(k)}\t^{(+1)^{\dag}}_{k,P(j)}+
\t^{(+1)^{\dag}}_{j,P(k)}\t^{(-1)^{\dag}}_{k,P(j)}\right]\prod_{i\neq 
j,k}\s^{\dag}_{i,P(i)}|0\ket
\label{id7}
\end{eqnarray}
where the permutations $P$ and $P'$ satisfy 
\begin{equation}
P'(k)=P(j),\;\;\;\;P'(j)=P(k),\;\;\;\;P'(i)=P(i)\;\;\forall i\neq j,k\;;
\end{equation}
hence $(-)^{P'}=-(-)^{P}$. Substituting Eq.(\ref{id7}) into 
Eq.(\ref{sq}) and taking into account that
\begin{equation}
\t^{(-1)^{\dag}}_{j,P(k)}\t^{(+1)^{\dag}}_{k,P(j)}+
\t^{(0)^{\dag}}_{j,P(j)}\t^{(0)^{\dag}}_{k,P(k)}+
\t^{(+1)^{\dag}}_{j,P(k)}\t^{(-1)^{\dag}}_{k,P(j)}=
\s^{\dag}_{j,P(j)}\s^{\dag}_{k,P(k)}
\end{equation}
one obtains
\begin{equation}
[..]=3\sqrt{2}(N-1)|\F_{AF}^{0,0}\ket+\sqrt{2}
\sum_{P}(-)^{P}\sum_{j}\sum_{k\neq j}
\s^{\dag}_{j,P(j)}\s^{\dag}_{k,P(k)}\prod_{i\neq 
j,k}\s^{\dag}_{i,P(i)}|0\ket=\sqrt{2}(N^{2}-1)|\F_{AF}^{0,0}\ket\;,
\end{equation}
that is, Eq.(\ref{ID4}).

}


\begin{center}

REFERENCES
\end{center}
\bigskip 
\begin{references}
\bibitem{liebwu} E. H. Lieb and F. Y. Wu, {\it Phys. Rev. Lett.} {\bf 
20}, 1445 (1968).
\bibitem{bethe} H. A. Bethe, {\it Zeits. f. Phys.} {\bf 71}, 205 (1931); E. 
H. Lieb and W. Liniger, {\it Phys. Rev.} {\bf 130}, 1605 (1963).
\bibitem{fermbethe} J. B. McGuire, {\it J. Math. Phys.} {\bf 6}, 432 (1965); 
M. Flicker and E. H. Lieb, {\it Phys. Rev.} {\bf 161}, 179 (1967).
\bibitem{yang2} C. N. 
Yang, {\it Phys. Rev. Lett.} {\bf 19}, 1312 (1967); M. Gaudin, {\it 
Phys. Letters} {\bf A 24}, 55 (1967).
\bi{suzhao} Gang Su, Bao-Heng Zhao and Mo-Liu Ge, {\it Phys. Rev.} 
{\bf B 46}, 14909 (1992).
\bi{ligruber} You-Quou Li and Christian Gruber, {\it Phys. Rev. Lett.} 
{\bf 80}, 1034 (1998).
\bi{houpeng} Boyu Hou, Dantao Peng and Ruihong Yue, {\tt 
cond-mat/0010485} (2000).
\bibitem{lieb} E. H. Lieb, {\it Phys. Rev. Lett.} {\bf 62}, 1201 (1989).
\bi{nagaoka} Y. Nagaoka {\it Phys. Rev.} {\bf 147}, 392 (1966).
\bi{mielke} A. Mielke {\it J. Phys} {\bf A 24}, L73 (1991); A. Mielke 
{\it J. Phys} {\bf A 25}, 4335 (1992); A. Mielke {\it Phys. Lett.} 
{\bf A 174}, 443 (1993); A. Mielke {\it Phys. Rev. Lett.} {\bf 82},  
4312 (1999); A. Mielke, {\it J. Phys} {\bf A 32}, 8411 (1999).
\bi{tasaki} H. Tasaki {\it Phys. Rev. Lett.} {\bf 69}, 1608 (1992); 
H. Tasaki {\it J. Stat. Phys.} {\bf 84}, 535 (1996).
\bi{wang} D. F. Wang, {\tt cond-mat/9604121}.
\bi{patterson} J. D. Patterson, {\it Phys. Rev.} {\bf B 6}, 1041 (1972).
\bi{vDV} P. van Dongen and D. Vollhardt, {\it Phys. Rev.} {\bf B 40}, 
7252 (1989).
\bi{verges} J. A. Verges, F. Guinea, J. Galan, P. van Dongen, G. 
Chiappe and E. Louis, {\it Phys. Rev.} {\bf B 49}, 15400 (1994).
\bi{brandt} U. Brandt and A. Giesekus, {\it Phys. Rev. Lett.} {\bf 
68}, 2648 (1992).
\bi{mita} A. Mielke, {\it J. Phys.} {\bf A 24}, 3311 (1991); A. 
Mielke and H. Tasaki, {\it Commun. Math. Phys.} {\bf 158}, 341 (1993). 
\bibitem{pseudo}  O.J. Heilmann and E. H. Lieb, 
{\it Annals N.Y. Acad. Sci.} {\bf 172}, 583 (1971).
\bibitem{ya} Cheng Ning Yang, {\it Phys. Rev. Lett.} {\bf 63}, 2144 (1989).
\bibitem{yaza} Cheng Ning Yang and S. C. Zhang, {\it Mod. Phys. Lett.} {\bf 
B 4}, 759 (1990).   
\bibitem{ssc2001} Michele Cini and Gianluca Stefanucci, {\it Solid State 
Communications} {\bf 117}, 451 (2001).
\bibitem{jop2001} Michele Cini and Gianluca Stefanucci, {\it J. Phys.: 
Condens. Matter} {\bf 13}, 1279 (2001).
\bibitem{jop2002} Gianluca Stefanucci and Michele Cini, {\it J. Phys.: 
Condens. Matter} {\bf 14}, 2653 (2002).
\bibitem{liebmattis} E. H. Lieb and D. C. Mattis, {\it J. Math. Phys.} {\bf 
3}, 749 (1962).
\bibitem{gutz} M. C. Gutzwiller, {\it Phys. Rev. Lett.} {\bf 10}, 
159 (1963).
\bibitem{sqt} S. Q. Shen, Z. M. Qiu and G. S. Tian, Phys. Rev. 
Lett. {\bf 72}, 1280 (1994).

\end{references}
\end{document}